\documentclass[iop]{emulateapj}
\usepackage{graphicx}
\usepackage{multirow}
\usepackage{array}
\usepackage{subfigure}
\usepackage{hyperref}
\usepackage{microtype}
\usepackage[usenames]{color}

\begin{document}
\newcommand{\beq}{\begin{equation}}
\newcommand{\eeq}{\end{equation}}

\newcommand{\beqa}{\begin{eqnarray}}
\newcommand{\eeqa}{\end{eqnarray}}

\newcommand{\avg}[1]{\langle{#1}\rangle}

\newcommand{\Msun}{M_{\rm\odot}}
\newcommand{\hiMsun}{h^{-1}  M_{\rm\odot}}


\newcommand{\hiGpc}{h^{-1}{\rm Gpc}}
\newcommand{\hiMpc}{h^{-1}{\rm Mpc}}
\newcommand{\hikpc}{h^{-1}{\rm kpc}}

\newcommand{\Mvir}{M_\mathrm{vir}}
\newcommand{\Rvir}{R_\mathrm{vir}}
\newcommand{\Deltavir}{\Delta_\mathrm{vir}}
\newcommand{\kms}{{\rm s^{-1}km}}

\newcommand{\rmax}{r_\mathrm{max}}
\newcommand{\vmax}{\mathrm{v}_\mathrm{max}}
\newcommand{\vnow}{\mathrm{v}_\mathrm{0}}
\newcommand{\vtoday}{\mathrm{v}_\mathrm{0}}
\newcommand{\vacc}{\mathrm{v}_\mathrm{ac}}
\newcommand{\vpeak}{\mathrm{v}_\mathrm{pk}}
\newcommand{\vcut}{\mathrm{v}_\mathrm{cut}}
\newcommand{\Macc}{M_\mathrm{ac}}
\newcommand{\Mtoday}{M_\mathrm{0}}
\newcommand{\Mpeak}{M_\mathrm{pk}}

\newcommand{\zhalf}{z_{\rm 1/2}}
\newcommand{\Cdyn}{C_{\rm dyn}}

\journalinfo{The Astrophysical Journal, {\rm 763:70 (18pp), 2013 February 1}}
\submitted{Received 2012 September 14; accepted 2012 November 28; published
2013 January 9}
\shortauthors{Wu et al.}
\shorttitle{Rhapsody Simulations of Massive Galaxy Clusters}

\author{Hao-Yi Wu$,^{1,2}$  Oliver Hahn,$^1$  Risa H. Wechsler,$^1$
Yao-Yuan Mao,$^1$ and Peter S. Behroozi$^1$
}
\affil{ 
$^1$Kavli Institute for Particle Astrophysics and Cosmology;
 Physics Department, Stanford University, Stanford, CA 94305, USA\\
 SLAC National Accelerator Laboratory, Menlo Park, CA 94025, USA\\
$^2$Physics Department, University of Michigan, Ann
Arbor, MI 48109, USA; {\tt hywu@umich.edu}}

\title{Rhapsody. I. Structural Properties and Formation History From a Statistical Sample of Re-simulated Cluster-size Halos}

\begin{abstract}
  We present the first results from the {\sc Rhapsody} cluster
  re-simulation project: a sample of 96 ``zoom-in'' simulations of
  dark matter halos of $10^{14.8\pm 0.05}\hiMsun$, selected from a 1
  ${h^{-3}}{\rm Gpc}^3$ volume.  This simulation suite is the first to
  resolve this many halos with $\sim 5\times10^6$ particles per halo
  in the cluster mass regime, allowing us to statistically
  characterize the distribution of and correlation between halo
  properties at fixed mass.  We focus on the properties of the main
  halos and how they are affected by formation history, which we
  track back to $z=12$, over five decades in mass.  We give particular
  attention to the impact of the formation history on the density
  profiles of the halos.  We find that the deviations from the
  Navarro--Frenk--White (NFW) model and the Einasto model depend on
  formation time.  Late-forming halos tend to have considerable
  deviations from both models, partly due to the presence of massive
  subhalos, while early-forming halos deviate less but still
  significantly from the NFW model and are better described by the
  Einasto model.  We find that the halo shapes depend only moderately
  on formation time. Departure from spherical symmetry impacts the
  density profiles through the anisotropic distribution of massive
  subhalos.  Further evidence of the impact of subhalos is provided by
  analyzing the phase-space structure.  A detailed analysis of the
  properties of the subhalo population in {\sc Rhapsody} is presented
  in a companion paper.
\end{abstract}

\section{Introduction}

Galaxy clusters are powerful probes of cosmological parameters and
have played a key role in the development of the current $\Lambda$CDM
paradigm (see, e.g., \citealt{Allen11} for a review).  For example, the
spatial distribution and abundance of galaxy clusters reflect the
growth rate of large-scale structure and the expansion rate of the
universe, providing constraints on dark matter and dark energy
\citep[e.g.,][]{Vikhlinin09,Mantz10, Rozo10}, neutrino mass
\citep[e.g.,][]{Mantz10b, Reid10}, and the validity of general
relativity on cosmic scales \citep[e.g.,][]{Rapetti10, Rapetti12}.  In
the near future, the massive influx of multi-wavelength data (e.g.,
SPT,\footnote{The South Pole Telescope; http://pole.uchicago.edu/}
ACT,\footnote{Atacama Cosmology Telescope; \nolinebreak
  http://www.princeton.edu/act/ }
Planck,\footnote{http://www.esa.int/planck} 
eRosita,\footnote{Extended
  ROentgen Survey with an Imaging Telescope Array;
  http://www.mpe.mpg.de/eROSITA} 
PanSTARRS,\footnote{The Panoramic
  Survey Telescope \& Rapid Response System;
  http://pan-starrs.ifa.hawaii.edu/} 
DES,\footnote{The Dark Energy Survey; http://www.darkenergysurvey.org/}
Euclid,\footnote{http://sci.esa.int/euclid/} 
LSST\footnote{The Large
  Synoptic Survey Telescope; http://www.lsst.org/}) will greatly
enhance the sample size of galaxy clusters and reduce the statistical
uncertainties in cluster cosmology.  However, the constraining power
of galaxy clusters will depend on how well various systematic
uncertainties can be controlled, including the relations between
observable properties and mass \citep[e.g.,][]{Rozo12}; the robustness
of cluster identification and centering \citep[e.g.,][]{Rykoff12}; and
the effect of viewing angle and projection \citep[e.g.,][]{WhiteM10}.

One essential way to understand these systematic uncertainties is
through $N$-body simulations, which have been applied to study galaxy
clusters for more than a decade (e.g.,
\citealt{Tormen97,Moore98,Ghigna98}; also see
\citealt{KravtsovBorgani12} for a more general review).  In the era of
large-sky survey and precision cosmology, it is desirable to have
controlled simulation samples that can help us understand the {\em
statistical distribution} of the properties of galaxy clusters and the
{\em correlation} between observables, as well as their detailed
structures and evolution.  Since massive galaxy clusters are rare,
cosmological simulations need to cover a large volume to include a
fair number of these systems (e.g., the MultiDark simulation
(\citealt{Prada11}) and the recent Millennium XXL simulation
(\citealt{Angulo12})).  However, given limited computational
resources, the detailed substructures of halos are not well-resolved
in these simulations.  Instead of using a cosmological volume with a
single resolution, one can focus on particular systems and re-simulate
them with higher resolution.  This so-called ``zoom-in'' technique
provides a powerful way to study individual cluster systems in detail
in a cosmological context
\citep[e.g.,][]{Tormen97,Moore99,Navarro04,Gao05b,Reed05}.
Nevertheless, so far most zoom-in simulations have focused only on a
small number of cluster-size systems(e.g., the current high-resolution
Phoenix simulation (\citealt{Gao12}) ) and galactic halos (e.g., the
Via Lactea II simulation (\citealt{Diemand08}) and the Aquarius
simulations (\citealt{Springel08})).  Therefore, few statements have
been made about the statistical properties of well-resolved subhalos
in the mass regime of galaxy clusters\footnote{ We note that the
recent Marenostrum-MultiDark SImulations of galaxy Clusters
\citep{Sembolini12} present the largest set of hydrodynamical cluster
re-simulations to date.}.

In this work, we perform re-simulations of a large number of
cluster-forming regions in a cosmological volume (side length $1\
\hiGpc$) to create a high-resolution statistical cluster sample, {\sc
Rhapsody}, which stands for ``Re-simulated HAlo Population for
Statistical Observable--mass Distribution studY''.  The current sample
includes 96 halos of mass $10^{14.8\pm 0.05} \hiMsun$ with mass
resolution $1.3\times10^8 \hiMsun$.  One of the main goals of {\sc
Rhapsody} is to create a sample of cluster-size halos at fixed mass
that enables us to make statistical statements about the halo
population that is relevant for current and imminent cluster surveys.

\begin{figure} 
\includegraphics[width=\columnwidth]{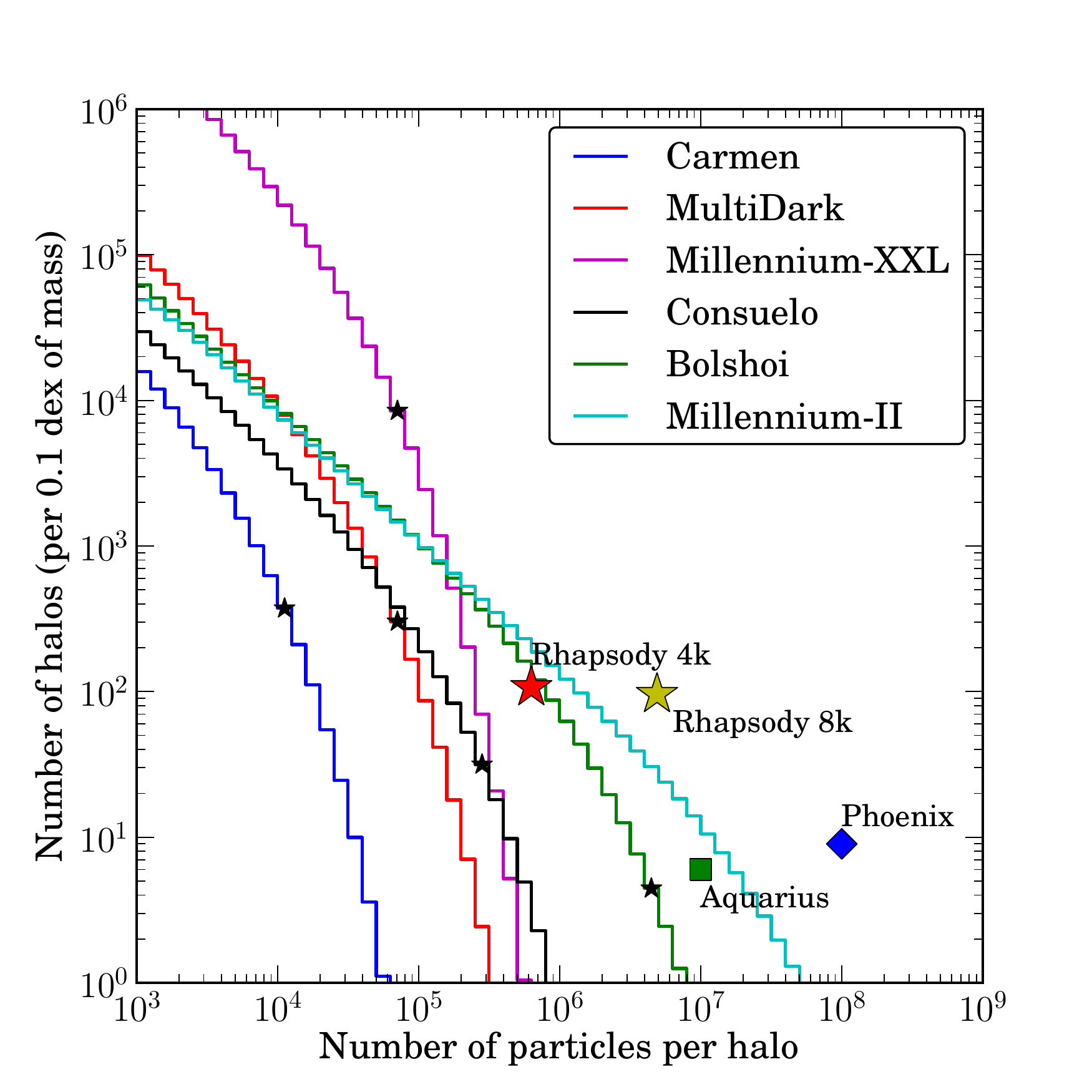}
\caption{Comparison of the halo samples in various $N$-body simulations;
{\sc Rhapsody} is in a unique statistical regime of well-resolved
massive halos.  The number of halos (per 0.1 dex in mass) is shown as
a function of number of particles inside the virial radius of the
halo.  Symbols correspond to halos in re-simulation projects; the {\sc
Rhapsody} 4K and 8K samples are shown as two colored stars ($\Mvir =
10^{14.8\pm 0.05} \hiMsun$).  Curves correspond to halos in different
cosmological volumes, and black stars on these curves correspond to
the number of halos of the same mass as {\sc Rhapsody}.  We note that
Consuelo and Carmen both include 50 volumes, and only one volume is
presented here.}
\label{fig:simcomp}
\end{figure}

In Figure~\ref{fig:simcomp}, we compare {\sc Rhapsody 8K} (main
sample) and {\sc Rhapsody 4K} (a factor of 8 lower in mass resolution)
to several $N$-body simulations in the literature (Millennium II
(\citealt{Boylan-Kolchin09}); Millenium XXL (\citealt{Angulo12});
Bolshoi (\citealt{Klypin10}); MultiDark (\citealt{Prada11}); Consuelo
and Carmen (from LasDamas; McBride et al. in preparation); Phoenix
(\citealt{Gao12}); Aquarius (\citealt{Springel08})).  Halos from
zoom-in simulations are presented by symbols, while halo populations
inside cosmological volumes are presented by curves with the shape of
the halo mass function.  {\sc Rhapsody} is in a unique regime in terms
of the number of halos in a narrow mass bin simulated with high
particle number.  It is also worth noting that {\sc Rhapsody} is
currently the largest sample of halos with more than a few times
$10^6$ particles per halo at any given mass in the literature.  Our
repeated implementation of the re-simulation method makes the
simulation suite statistically interesting and computationally
feasible.

The {\sc Rhapsody} sample is highly relevant to several current
observational programs.  For example, the galaxy cluster catalogs from
the SDSS \citep{Koester07,Wen09,Hao10} include many tens of thousands of
photometrically-selected clusters and have provided a rich sample for
multi-wavelength mass calibration \citep{Rozo09Richness,Rykoff12}, 
cosmological constraints \citep{Rozo07b,Rozo10}, and
studies of the cluster galaxy populations \citep{Hansen09}.  The
Cluster Lensing And Supernova survey with Hubble Multi-Cycle Treasury
Program (CLASH; \citealt{Postman12}) focuses on 25 massive clusters
and aims to establish unbiased measurements of cluster
mass--concentration relation of these clusters.  Recently, \cite{vdLinden12}
published accurate weak lensing mass calibrations of 51
massive clusters, focusing on understanding various systematics for
cluster count experiments.  In addition, various X-ray programs have
been efficiently identifying massive clusters; for example, the ROSAT
Brightest Cluster Sample \citep{Ebeling00}, the ROSAT-ESO Flux-Limited
X-ray sample \citep{Bohringer04}, and the MAssive Cluster Survey
\citep{Ebeling10}.  These samples have achieved high completeness and
provided a relatively unbiased selection.  Relatively recently, massive
galaxy clusters have also been detected through the
Sunyaev--Zel'dovich (SZ) effect by ACT \citep{Marriage11}, SPT
\citep{Williamson11}, and Planck \citep{PlanckCluster11}.  These
detections have
ushered in an era of high-purity detection of high-redshift galaxy
clusters.  The {\sc Rhapsody} sample is in a mass regime similar to
these observational programs and can provide a statistical description
of the dark matter halos associated with these clusters.

This paper presents the first results from the {\sc Rhapsody}
simulations.  We first characterize the formation history and the
density profiles of the 96 main halos.  We then explore how formation
history impacts halo concentration and the deviation from the
Navarro--Frenk--White (NFW) profile.  We find that the deviations from
the NFW model systematically depend on formation time and are impacted
by the presence of massive subhalos.

We connect the density profile to the phase-space structures of halos
and find that late-forming halos tend to have outflows within $\Rvir$,
which can also be attributed to massive subhalos.  We have also
investigated the shape parameters for the spatial distributions and
velocities of dark matter particles.  We find that the shape parameters,
after removing massive subhalos, have no strong correlation with
formation time, indicating that the deviation from spherical symmetry alone
cannot account for the trend of profile and formation time.

This paper is organized as follows.  In Section \ref{sec:sims}, we detail
the simulations.  We present the formation history of the main halos
in Section \ref{sec:MAH} and merger rate in Section \ref{sec:merger}.  In
Section \ref{sec:concentration_fit}, we present the density profiles and
compare various fitting functions; in Section \ref{sec:c_zhalf}, we
demonstrate the effect of formation history on the density profile.
In Section \ref{sec:Vr}, we analyze the impact of formation time on the
phase-space structure.  The shapes of spatial distributions and
velocities of dark matter particles and their alignments are discussed
in Section \ref{sec:shape}.  We conclude in Section \ref{sec:summary}.

In a second paper in this series \citep[][hereafter Paper II]{Wu12b},
we will present the properties of subhalos in our sample and the
impact of formation time on them, which is more complex in the
sense that the subhalo properties depend on the selection method of
subhalos, the stripping experienced by a subhalo, and the resolution
of the simulation.

\section{The Simulations}\label{sec:sims}
\begin{table*}
\centering
\begin{tabular}{
>{\centering}p{2cm}<{\centering}  
>{\centering}p{2.5cm}<{\centering} 
>{\centering}p{2.5cm}<{\centering} 
>{\centering}p{2.5cm}<{\centering} 
>{\centering}p{3cm}<{\centering} 
c
}
\hline
{Type} &
{Name} &
{Mass Resolution} &
{Force Resolution} & 
{Number of Particles}&
{Number of Particles}\\ 
{} &
{} &
{[$\hiMsun$]} &   
{[$\hikpc$]} &
{in Simulation} &
{in Each Targeted Halo} \\
\hline
\hline
\multirow{1}{*}{Full Volume} 
& {\sc Carmen}   & 4.94$\times 10^{10}$ & 25 
&   $1120^3$ 
& 12K \\ 
\hline
\multirow{2}{*}{Zoom-in} 
& 
{\sc Rhapsody} 4K  & 
1.0$\times 10^{9}$ & 
6.7 & 
5.4M$^a$ /$4096^3 (equiv.)$ & 
0.63M$^b$\\  
& 
{\sc Rhapsody} 8K  &
1.3$\times10^{8}$ & 
3.3 & 
42M$^a$ / $8192^3 (equiv.)$ & 
4.9M$^b$\\  
\hline
\multicolumn{6}{l}{$^a$The mean number of high-resolution particles in each zoom-in region.}  \\
\multicolumn{6}{l}{$^b$The mean number of high-resolution particles within the $\Rvir$ of each targeted halo.} \\
\end{tabular}
\caption{Simulation parameters. }
\label{tab:sims}
\end{table*}

The {\sc Rhapsody} sample includes 96 cluster-size halos of mass
$\Mvir = 10^{14.8 \pm 0.05} \hiMsun$, re-simulated from a cosmological
volume of 1 $\hiGpc$.  Each halo was simulated at two resolutions:
$1.3\times10^{8}\hiMsun$ (equivalent to $8192^3$ particles in this
volume), which we refer to as ``{\sc Rhapsody 8K}'' or simply ``{\sc
  Rhapsody}''; and $1.0\times10^9 \hiMsun$ (equivalent to $4096^3$
particles in this volume), which we refer to as ``{\sc Rhapsody 4K}.''
These two sets allow detailed studies of the impact of resolution. The
simulation parameters are summarized in Table~\ref{tab:sims}.

The initial conditions were generated with the multi-scale initial
condition generator {\sc Music} \citep{HahnAbel11}.  The particles were
then evolved using the public version of {\sc Gadget-2}
\citep{Springel05}.  The halo finding was performed with the
phase-space halo finder {\sc Rockstar} \citep{Behroozi11rs}.  Finally,
merger trees were constructed with the gravitationally-consistent code
of \cite{Behroozi11tree}.  We provide more details on our methods
below.

All simulations in this work are based on a $\Lambda$CDM cosmology
with density parameters $\Omega_m = 0.25$, $\Omega_\Lambda= 0.75$,
$\Omega_b = 0.04$, spectral index $n_s = 1$, normalization $\sigma_8 =
0.8$, and Hubble parameter $h=0.7$.

\begin{figure*}
\centering
 \vspace{0.2in}
\includegraphics[width=7in]{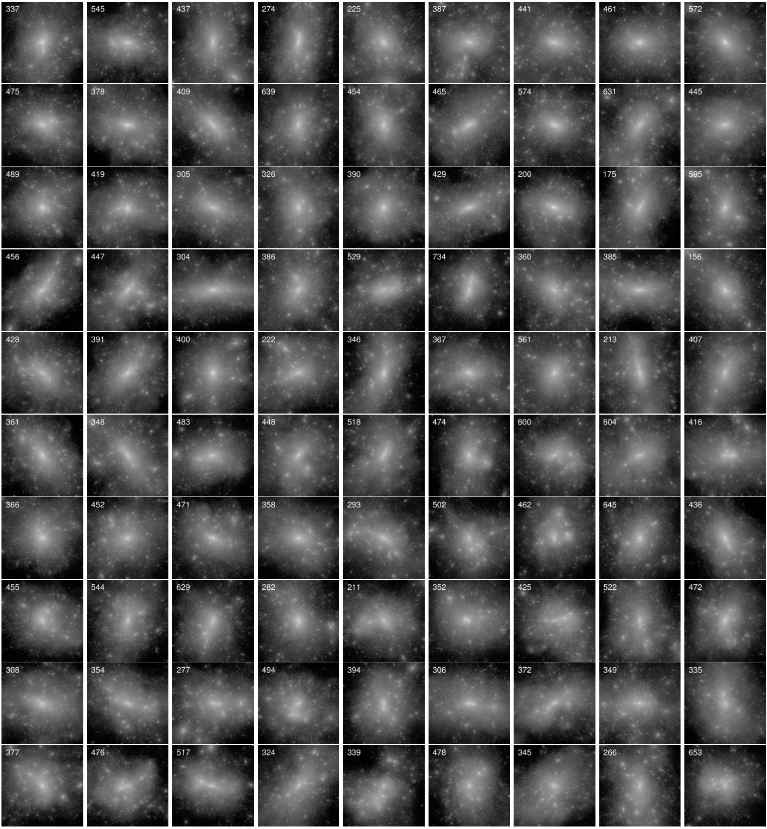}
 \vspace{0.2in}
\caption{Images of 90 {\sc Rhapsody} halos at $z=0$.  The halos are first
sorted by concentration (high concentration on the upper rows, low on the bottom).  In
each row, the halos are then sorted by the number of subhalos
(selected with $\vmax>100 \ \kms$, high number of subhalos on
the left columns, low on the right).  Each image has a physical extent of 4 $\hiMpc$ on
a side, which is slightly larger than the average virial radius of 1.8 $\hiMpc$.
}
\label{fig:thumbnails}
\end{figure*}

\begin{figure*}[t]
\centering
\includegraphics[width=7in]{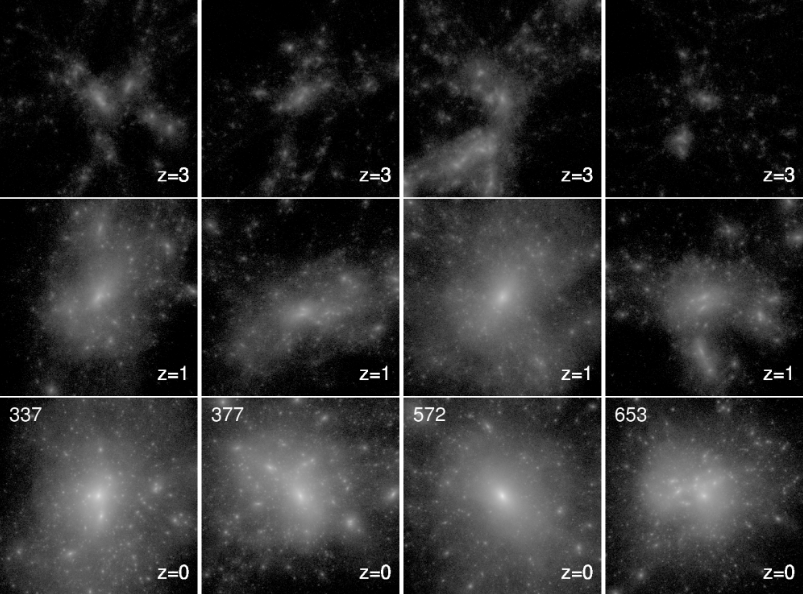}
\caption{Evolution of four {\sc Rhapsody} halos.  From top to bottom, the
 images show the progenitors of four halos at $z=3$ and $z=1$, and the
  halo at $z=0$.  The four halos chosen are the corners of Figure 2.  From
  left to right, they have high concentration, high subhalo number [337];
  low concentration, high subhalo number [377]; high concentration, low
  subhalo number [572]; low concentration, low subhalo number [653].
   Halo 572 has the highest concentration, the least late-time
   accretion, and the most dominant central halo of our full sample.  It is
   also the halo with the most massive progenitor at $z=3$. Each panel has a comoving
   extent of 4 $\hiMpc$ on a side, and is centered on the most massive progenitor in each case. }
\label{fig:haloevolutionpix}
\end{figure*}

Figure \ref{fig:thumbnails} shows the images of 90 halos at $z=0$ in
the {\sc Rhapsody 8K} sample.  Halos are sorted by their concentration
and subhalo number, as described in the following sections.  Figure
\ref{fig:haloevolutionpix} shows the evolution of 4 individual halos,
selected as extremes in the distribution of concentration and subhalo
number.  Movies and images for each individual halo are available at
{\tt http://risa.stanford.edu/rhapsody/}.

\subsection{The cosmological volume}

Our re-simulations are based on one of the {\sc Carmen} simulations
from the LArge Suite of DArk MAtter Simulations ({\sc LasDamas};
McBride et al.).  The simulation represents a cosmological volume of 1
$\hiGpc$ with $1120^3$ particles.  Its initial conditions are
generated with the code of \cite{Crocce06} based on the second-order
Lagrangian perturbation theory (2LPT), and the $N$-body simulation was
run with the {\sc Gadget-2} code.  {\sc Rhapsody} uses the same
cosmological parameters as {\sc Carmen}.

When selecting targets for re-simulation from the massive end of the
halo mass function, we choose a mass bin that is narrow enough so that
mass trends of halo properties are negligible but at the same time
wide enough to include a sufficient number of halos for statistical
analyses.  Here we focus on a 0.1 dex bin surrounding $\rm
log_{10}\Mvir = 14.8$.  This mass range allows us to select $\sim 100$
halos in a narrow mass range, and is well-matched to the masses of the
massive clusters studied in X-ray, SZ, and optical cluster surveys.

\subsection{Initial conditions}

For each of the halos in our sample, we generate multi-resolution
initial conditions using the {\sc Music} code \citep{HahnAbel11}.  We
use the same white noise field of {\sc Carmen} ($1024^3$ of its
$1120^3$ modes) to generate the large-scale perturbations consistent with
{\sc Carmen}.  The equivalent resolution ranges from $256^3$ in the
lowest resolution region to $8192^3$ ($4096^3$ for the 4K sample) in
the highest resolution region.  In between, the mass resolution
changes by factors of 8 every 8 times the mean inter-particle
distance.

For each of our re-simulation targets, we choose a zoom-in volume that
is 40\% larger than the Lagrangian volume of the friends-of-friends
halo at $z = 0$.  This choice has been tested to provide a
well-converged dark matter density profile in our convergence tests.
With this setting, no low-resolution particle was found within the
virial radius of any targeted halo.  The typical number of high
resolution particles per simulation is thus about 42/5.4 million for
8K/4K with a standard deviation of $18\%$.

In {\sc Music}, particle displacements and velocities have been
computed from the multi-scale density field using 2LPT at a starting
redshift of 49, in accordance with {\sc Carmen}.  The use of 2LPT is
important for statistical studies of such massive systems since their
masses depend on the accuracy of the initial conditions (e.g.,
\citealt{Crocce06,Tinker08,Reed12,Behroozi12}; and McBride et al., in
preparation).

\subsection{Gravitational evolution}
After generating the initial conditions, we evolve each
cluster-forming region using the public version of the {\sc Gadget-2}
code \citep{Springel05}.  Gravitational forces are computed using two
levels of particle-mesh together with the force tree to achieve a
force resolution of comoving 3.3/6.7 $\hikpc$ in the {\sc Rhapsody}
8K/4K for particles in the high resolution region. 
For each simulation, we save 200/100 snapshots logarithmically spaced in scale
factor $a$ between $a=0.075$ and $a=1$ for the 8K/4K sample.

We note that the virial masses of the re-simulated halos change
somewhat with the improved resolution.  As a result, a fraction of the
halos fall outside the narrow targeted mass range $\rm log_{10}\Mvir =
14.8\pm 0.05$.  In most cases, the masses scatter slightly upwards.
We discard those halos falling outside the targeted mass bin of {\sc
  Rhapsody} to keep the mass selection clean.  In principle, to obtain
all halos in the $14.8\pm0.05$ mass bin in the re-simulated sample,
one needs to re-simulate a wider range of masses around $14.8$ to
include all halos that end up in the targeted bin.  However, the large
suite of re-simulations thus required is beyond the scope of this
work.  Thus, we note that {\sc Rhapsody} does not strictly include the
complete sample of halos within $\rm log_{10}\Mvir = 14.8\pm 0.05$ in
either the original volume or the re-simulations.  However, we do not
expect this fact to affect the results presented in this paper,
because the main approach in this paper is stacking all halos in {\sc
  Rhapsody} for sufficient statistics and our sample should be
unbiased.  Global statistics for halos in this bin in the entire
cosmological volume (for example, the two-point correlation function)
are not used in the present work.

\subsection{Halo and subhalo identification}

Our simulations are post-processed with the adaptive phase-space halo
finder {\sc Rockstar} \citep{Behroozi11rs}, which can achieve high
completeness in finding subhalos even in dense environments (
see also \citealt{Knebe11}).  Based on the phase-space information, small
subhalos passing through the dense central region of the main halo can
be robustly identified.  This feature is especially important for
studying the subhalo populations, which we focus on in Paper II.  We
note that the algorithm is only applied to high-resolution particles
in the simulations.

{\sc Rockstar} pays special attention to major merger events (two
halos of similar mass merge with each other), which arise frequently
in the formation history of {\sc Rhapsody} halos (because of their
high masses) and often cause difficulties in the construction of
merger trees.  During a major merger between two halos, a large
fraction of dark matter particles appear as unbound to either of the
merging halos, even though they are bound to the entire merging
system.  Therefore, regular unbinding procedures tend to result in
ambiguities or inconsistencies in halo mass assignment.  {\sc
  Rockstar} addresses this issue by computing the gravitational potential of
the entire merging system, thus making the mass evolution of halos
self-consistent across time steps.

\subsection{Merger trees}

We apply the gravitationally-consistent merger tree algorithm
developed by \cite{Behroozi11tree} to the 200/100 output snapshots
which were saved between $z = 12.3$ and $z = 0$ for {\sc Rhapsody}
8K/4K.  The idea behind this new merger tree implementation is that
the stochasticity in $N$-body simulations often leads to failures in
halo finding.  For example, the halo finder might find a spurious halo
that is in reality a random density fluctuation at a certain time
step, or the halo finder might miss a halo because it falls below the
detection threshold at that particular time step.  Given these
limitations in halo finders, previous implementations of merger trees
often encounter problems in linking halos across different time steps.
The gravitationally-consistent merger tree algorithm resolves this
issue by comparing adjacent time steps to recover missing subhalos and
remove spurious halos, thereby improving the completeness and purity
of the halo catalogs and ensuring correct linking of halos across time
steps.  This algorithm compares two adjacent time steps and can be
summarized as follows: (1) It takes the halos at the later time step
and evolves their positions and velocities backward in time, deciding
whether the progenitors are missing or incorrectly linked.  (2) It
takes the halos at the earlier time step and looks for its descendant
in the later time step.  If the descendant is missing, the algorithm
decides whether a merger occurs or the current halo is spurious.  For
details of the implementation, we refer the reader to
\cite{Behroozi11tree}.

\begin{table*}[t]
\centering
\begin{tabular}{ >{\centering}p{2cm}<{\centering} | 
>{\centering}p{2cm}<{\centering} 
>{\centering}p{2cm}<{\centering} 
|
>{\centering}p{2cm}<{\centering} 
>{\centering}p{2cm}<{\centering} 
 | c }
\hline&&&&\\ 
Property & Median & Frac. Scatter & Mean & Frac. SD  & Def. \\
&&&&\\\hline&&&&\\ 
$\Mvir\ [\hiMsun]$ & $6.4\times10^{14}$ & 0.072 & $6.4\times10^{14}$ & 0.061 &  \multirow{4}{*}{\S\ref{sec:mass}}\\
$\Rvir\ [\hiMpc]$ & 1.8 & 0.024 & 1.8 & 0.028 & \\
$\sigma_v\ [\kms]$ & 1,400 & 0.041 & 1,400 & 0.039 & \\
$\vmax\ [\kms]$ & 1,300 & 0.036 & 1,300 & 0.042 &\\
&&&&\\\hline&&&&\\ 
$M_{200m}$ & $7.3\times10^{14}$ & 0.083 & $7.3\times10^{14}$ & 0.077 & \multirow{4}{*}{$\hiMsun$; \S\ref{sec:mass}}\\
$M_{200c} $ & $5.9\times10^{14}$ & 0.065 & $5.9\times10^{14}$ & 0.056 & \\
$M_{500m} $ & $5.0\times10^{14}$ & 0.063 & $5.0\times10^{14}$ & 0.062 & \\
$M_{500c} $ & $3.4\times10^{14}$ & 0.13 & $3.4\times10^{14}$ & 0.12 & \\
&&&&\\\hline&&&&\\ 
$R_{200m} $ & 2.3 & 0.028 & 2.3 & 0.026 & \multirow{4}{*}{$\hiMpc$; \S\ref{sec:mass}}\\
$R_{200c} $ & 1.6 & 0.022 & 1.6 & 0.019 & \\
$R_{500m} $ & 1.3 & 0.021 & 1.3 & 0.021 & \\
$R_{500c} $ & 0.84 & 0.043 & 0.83 & 0.04 & \\
&&&&\\\hline&&&&\\ 
$z_{lmm}$ & 1.2 & 1.1 & 1.6 & 1.1 &   \multirow{4}{*}{\S\ref{sec:MAH}}\\
$z_{1/2}$ & 0.58 & 0.44 & 0.56 & 0.43 & \\
$z_{\alpha}$ & 0.67 & 0.058 & 0.67 & 0.059 & \\
$\gamma-\beta$ & 2.3 & 0.22 & 2.2 & 0.23 & \\
&&&&\\\hline&&&&\\ 
$r_s\ [\hiMpc]$ & 0.34 & 0.31 & 0.36 & 0.27 & \multirow{7}{*}{\S\ref{sec:concentration_fit}}\\
$c_{\rm NFW}$ & 5.3 & 0.26 & 5.3 & 0.23 & \\
$c_{\rm NFW-like}$ & 5 & 0.27 & 5.1 & 0.32 & \\
$c_{\rm Einasto}$ & 4.9 & 0.23 & 4.9 & 0.28 & \\
$\gamma_{\rm NFW-like}$ & 3.4 & 0.43 & 3.9 & 0.34 & \\
$\gamma_{\rm Einasto}$ & 3 & 0.13 & 3 & 0.14 & \\
$\alpha_{\rm Einasto}$ & 0.24 & 0.42 & 0.27 & 0.41 & \\
&&&&\\\hline&&&&\\ 
b/a & 0.75 & 0.12 & 0.76 & 0.12 &  \multirow{3}{*}{\S\ref{sec:PosShape}}\\
c/a & 0.63 & 0.12 & 0.63 & 0.12 & \\
T & 0.71 & 0.27 & 0.69 & 0.27 & \\
&&&&\\\hline&&&&\\ 
$b^{(v)}/a^{(v)}$ & 0.82 & 0.096 & 0.82 & 0.093 &   \multirow{3}{*}{\S\ref{sec:VelEllip}}\\
$c^{(v)}/a^{(v)}$ & 0.72 & 0.098 & 0.72 & 0.094 & \\
$\delta_{\sigma^2_{los}}$ & 0.17 & 0.32 & 0.18 & 0.3 & \\
&&&&\\\hline 
\end{tabular}

\caption[]{
Properties of {\sc Rhapsody} halos at $z=0$. The second column shows the median, the third
column corresponds to the ratio of the 68\% scatter to the median.  The fourth column shows
the sample mean and the fifth column corresponds to the ratio of the standard deviation to the
mean.}
\label{tab:properties}
\end{table*}

\subsection{Summary of halo properties}\label{sec:mass}

\begin{figure*}
\centering
\hspace*{-0.8in}
\includegraphics[width=2.5\columnwidth]{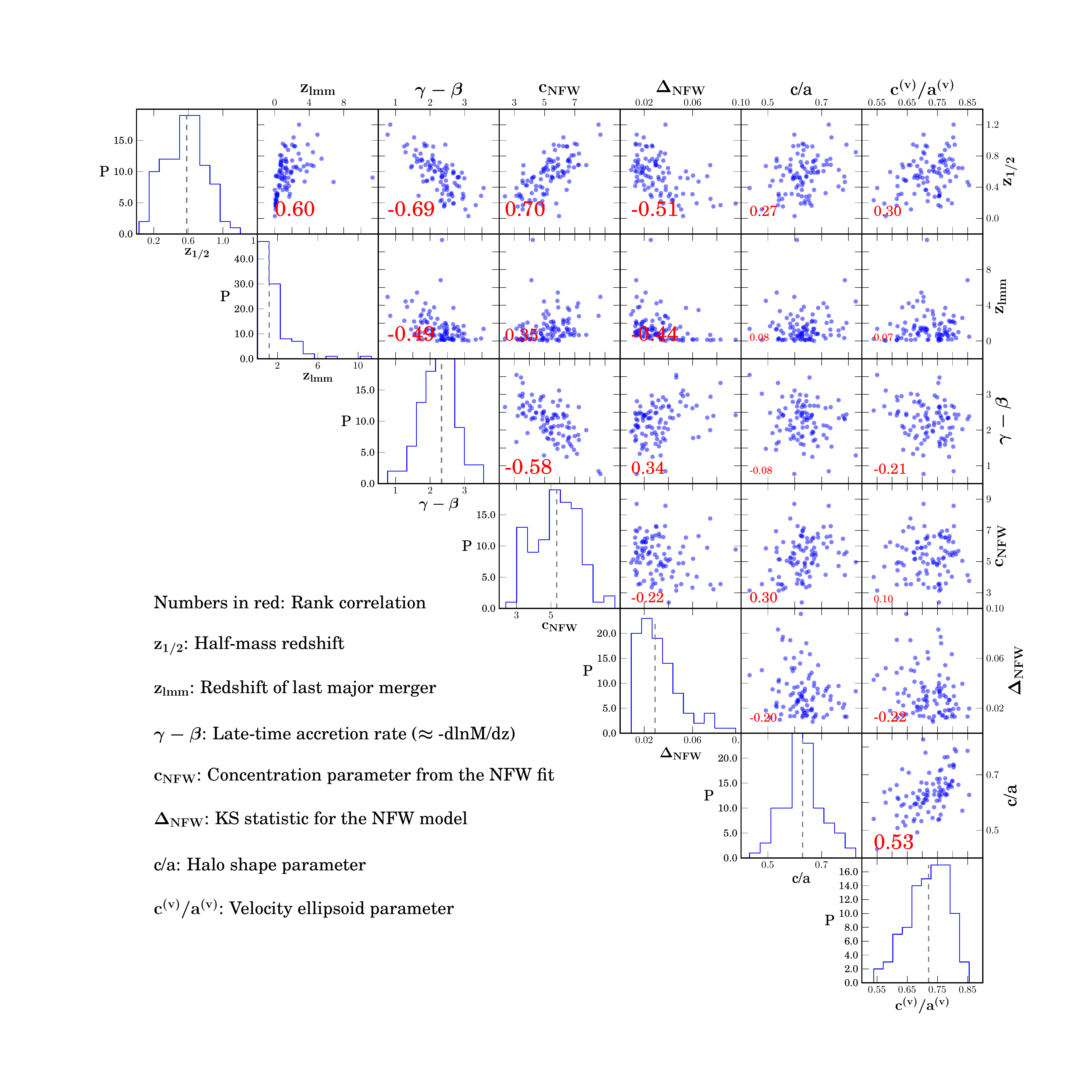}
\caption[]{Distributions of and correlations between main halo
properties and formation history parameters.  The number in red in
each panel shows the {\em rank correlation coefficient}, and its font
size reflects the magnitude of correlation.  }
\label{fig:cor}
\end{figure*}

In Table~\ref{tab:properties}, we summarize the key halo properties
discussed throughout this paper.  In Figure~\ref{fig:cor}, we present
the distributions of and correlations between several of these
properties.  We use the {\em rank correlations} throughout this work
to avoid the impact of outliers.

In this work, the halo mass definition is based on the spherical
overdensity of virialization, $\Deltavir$, with respect to the
critical density, $\rho_{\rm crit}$.  We use the center of the
phase-space density peak calculated by {\sc Rockstar} as the center of
a halo.  Based on this center, we draw a sphere with radius $\Rvir$ so
that the mean overdensity enclosed is equal to $\Deltavir\rho_{\rm
crit}$.  With the cosmological parameters used herein, $\Deltavir=94$
with respect to the critical density at $z=0$ \citep{BryanNorman98};
i.e., $\Deltavir =\Delta_{94c}$ =$ \Delta_{376m}$.  The subscripts $c$
and $m$ indicate the overdensities with respect to the critical
density and mean matter density.  For reference, in
Table~\ref{tab:properties} we list halo masses and radii based on
several commonly-used overdensity values: $\Delta_{200m} =
\Delta_{50c}$, $\Delta_{200c}$, $\Delta_{500m}=\Delta_{125c}$, and
$\Delta_{500c}$.

In Table~\ref{tab:properties}, we list two properties that are
closely related to halo mass: the maximum circular velocity and the
velocity dispersion of dark matter particles.  The maximum circular
velocities are defined at a radius $\rmax$ that maximizes $\sqrt{G
M(<r)/r}$:
\beq
\vmax =  \sqrt{\frac{G M(<\rmax)}{\rmax}}\ .
\label{eq:vmax}
\eeq
The velocity dispersion is calculated based on dark matter particles:
\beq
\sigma_{v}^2 =  \avg{|{\bf v - \bar{\bf v}}|^2} =
\frac{1}{N_p}\sum_{i=1}^{N_p} |{\bf v}_i - \bar{\bf v}|^2 \ .
\eeq
We note that the correlation between $\Mvir$ and $\sigma_{v}$ is 0.32,
indicating that there is a non-negligible residual mass--velocity
dispersion scaling despite the narrow mass range of our sample.  For
this reason, in the remainder of the paper, we have always verified
carefully that the quoted correlations between various properties are
not simply driven by mass.

\section{The buildup of cluster-size halos}

In this section, we present the mass accretion history and merger rate
of the main halos in {\sc Rhapsody}, paving the way for further
discussions of the impact of formation time on halo properties.  The
formation history of dark matter halos is known to correlate with
various halo properties, including their clustering, internal
structure, and subhalos \citep[e.g.,][]{Wechsler02, Zhao03, Harker06,
Hahn07, Maulbetsch07, Li08}.  On the other hand, the merger rate of
dark matter halos serves as a baseline for modeling several processes
in galaxy formation, including the build-up of stellar mass and
supermassive black holes, the star formation rate, the color and
morphology evolution
\citep[e.g.,][]{Kauffmann00,Hopkins06,DeLucia07,Behroozi12}.  In
addition, there has been an ongoing effort to measure merger rates in
observations \citep[e.g.,][]{Bell06,Lotz08,Xu12}.  While a study of
the implications for galaxy formation is beyond the scope of this
paper, we note that the merger rate provided here can be applied to
modeling galaxy formation in massive clusters.

\subsection{Mass accretion history}\label{sec:MAH}

\begin{figure}
\centering
\includegraphics[width=\columnwidth]{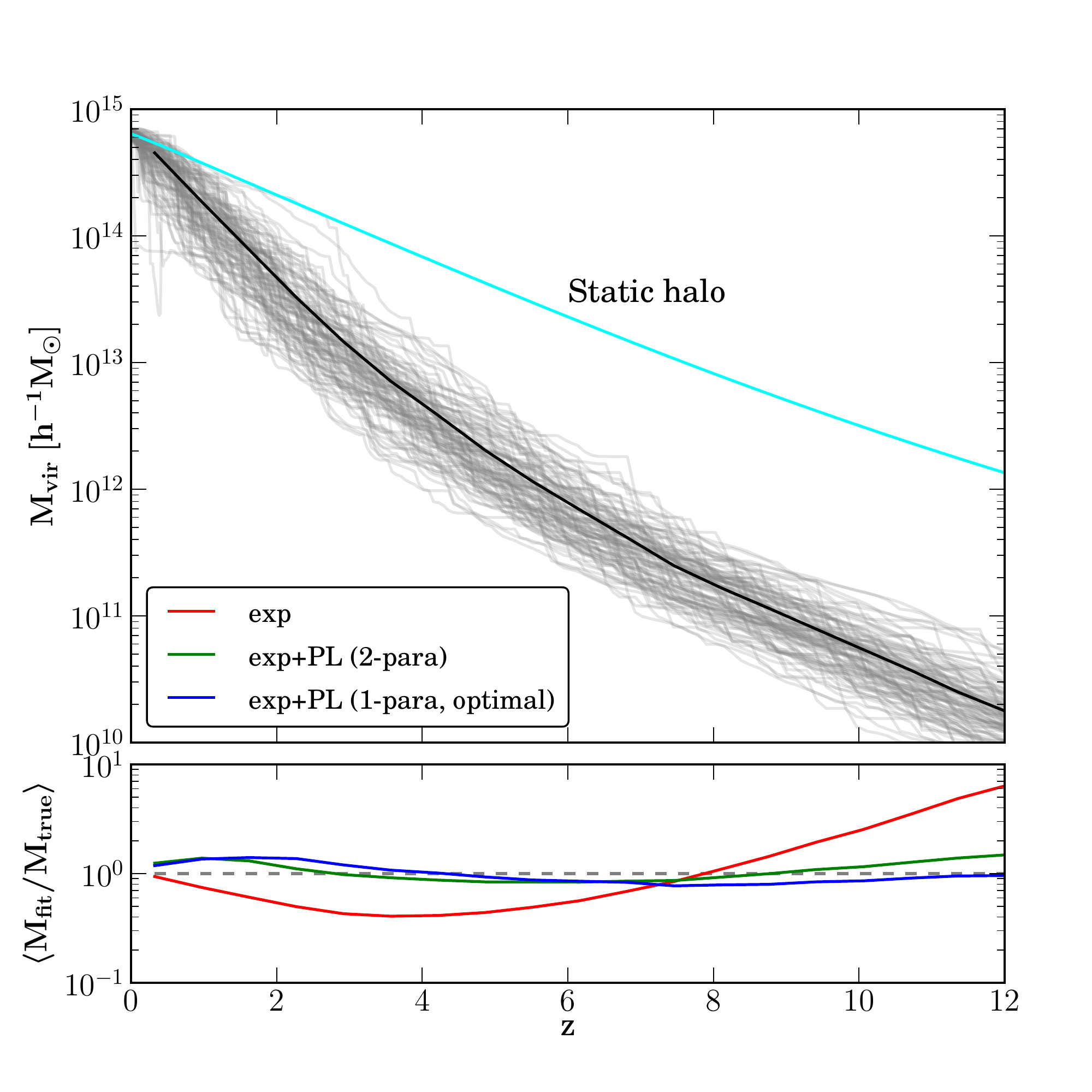}
\caption[]{Mass accretion history of the main halos in {\sc Rhapsody}.
  The gray curves indicate the trajectories of individual halos; the
  black curve indicates the average over all halos.  The cyan curve indicates
  the pseudo-evolution of a static halo.  We show the virial mass of
  the most massive progenitor at each output redshift.  The bottom
  panel shows the residual of the fit for three analytical models.\vspace{5pt}  }
\label{fig:MAH}
\end{figure}

Figure~\ref{fig:MAH} presents the mass evolution of the main halos in
{\sc Rhapsody}.  For each main halo identified at $z=0$, we search
through its merger tree to find the most massive progenitor at each
redshift.  The gray curves show the evolution of $\Mvir$ for
individual halos, and the black curve shows the average of all halos.
We note that the fractional dispersion is roughly constant for all
redshifts.  In the upper panel, we add a cyan line showing the pseudo
evolution of a static halo (with a non-evolving density profile and
$\Mvir=10^{14.8}\hiMsun$ at $z=0$) expected simply from the evolution
of $\Deltavir$ and the critical density \citep[e.g.,][]{Diemer12}.

Our halo formation histories span almost 5 orders of magnitude in mass
and cover the $13.8\,{\rm Gyr}$ between $z=12$ and $z=0$, allowing us
to test the validity of parameterizations proposed in the literature
\citep[e.g.,][]{Wechsler02, vdBosch02, Tasitsiomi04, Zhao09,
McBride09} over much larger ranges in both time and mass.  We fit the
following three models:
\begin{enumerate}
\item An exponential model \citep{Wechsler02}
\beq
M(z) = \Mtoday e^{-\alpha z}  \ .
\label{eq:MAH-1}
\eeq
We note that this model assumes a constant mass accretion rate
represented by the exponential growth index
\beq
-\frac{d\ln M}{dz} = \alpha \ .
\eeq
The curves in Figure~\ref{fig:MAH} do not follow straight lines, indicating a systematic deviation
from the pure exponential model.
For this model, a formation time proxy can be defined as
\beq
z_{\alpha} = \ln 2/\alpha  \ .
\label{eq:zalpha}
\eeq

\item An exponential-plus-power law model with two parameters \citep{McBride09}
\beq
M(z) = \Mtoday (1+z)^\beta e^{-\gamma z}\ . 
\label{eq:MAH-2}
\eeq
We note that 
\beq
-\frac{d\ln M}{dz} \approx  \gamma-\beta \ \ \mbox{when $z<<1$}\ .
\eeq
Thus, $\gamma-\beta$ can be used as a measure for the late-time
accretion rate.  
Analogous to Equation (\ref{eq:zalpha}), a formation time proxy can be
defined as
\beq
z_{\gamma-\beta} =  \ln 2/(\gamma-\beta)\ .
\eeq
One can alternatively solve $M(z_{\beta\gamma}) = \Mtoday/2$ numerically.  However, we note that
$z_{\beta\gamma}$, $z_{\gamma-\beta}$, and $\gamma-\beta$ are
completely correlated with each other.

\item An exponential-plus-power law model with one parameter,
motivated by Equation (\ref{eq:MAH-2}).  When fitting for
Equation (\ref{eq:MAH-2}), we observe that the two parameters $\beta$
and $\gamma$ are highly correlated (see Appendix \ref{app:MAH}).  We
thus adopt a 1-parameter model that incorporates this correlation
\beq
M(z) = \Mtoday(1+z) ^{-4.61+5.27 \gamma} e^{-\gamma z} \label{eq:MAH-3}.
\eeq
The two numerical coefficients are obtained by an optimization scheme
described in Appendix \ref{app:MAH}.
\end{enumerate}

Our fitting procedure and the goodness-of-fit are also discussed in
Appendix \ref{app:MAH}.  The bottom panel of Figure~\ref{fig:MAH}
shows the average of the residual for the individual fit, $\avg{M_{\rm
fit}/M_{\rm true}}$.  The pure exponential model does not provide the
curvature needed to fit the data.  The two exponential-plus-power law
models work almost equally well, but significant residual remains for
$z<2$.

\subsection{Merger rate}\label{sec:merger}
\begin{figure*}
\includegraphics[width=\columnwidth]{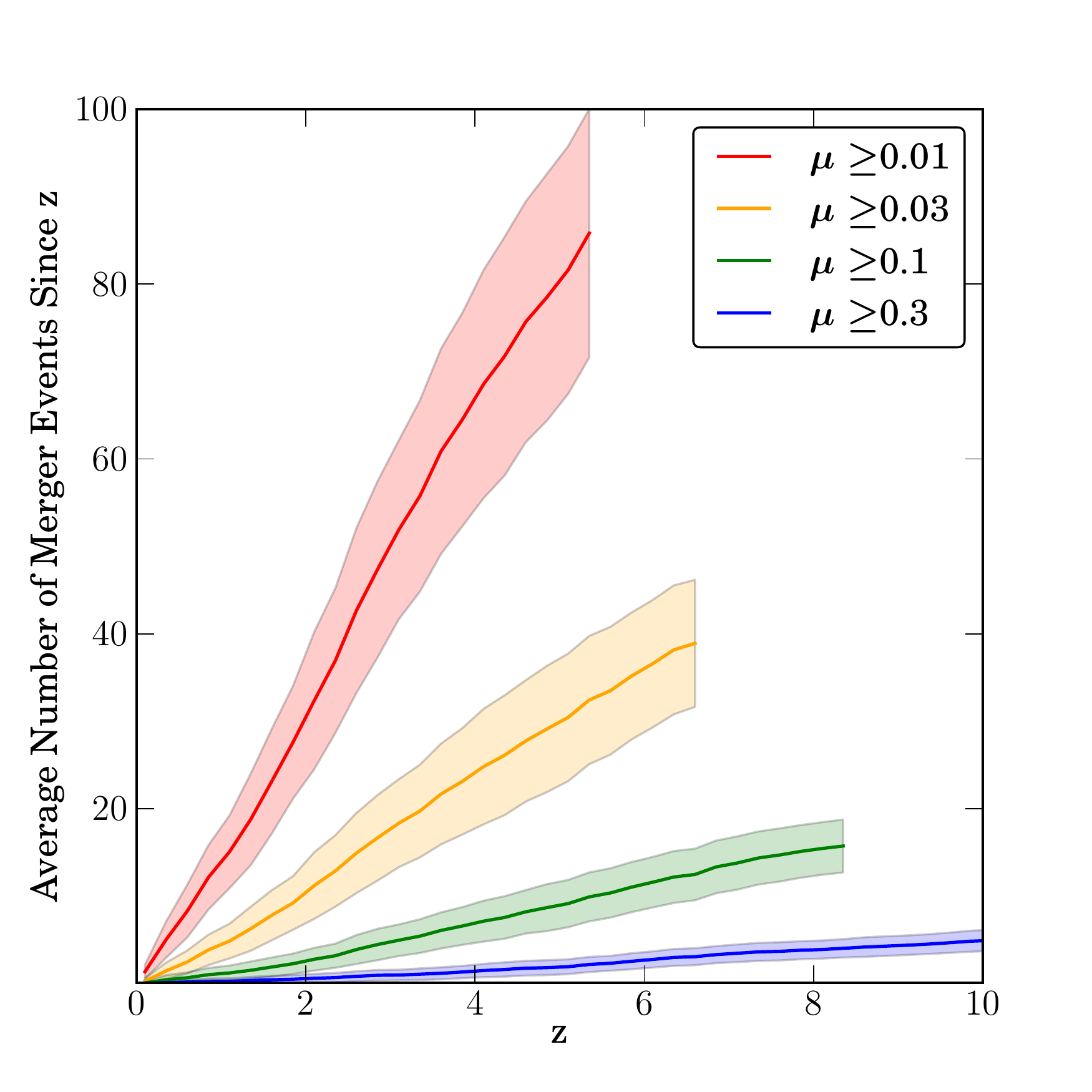}
\includegraphics[width=\columnwidth]{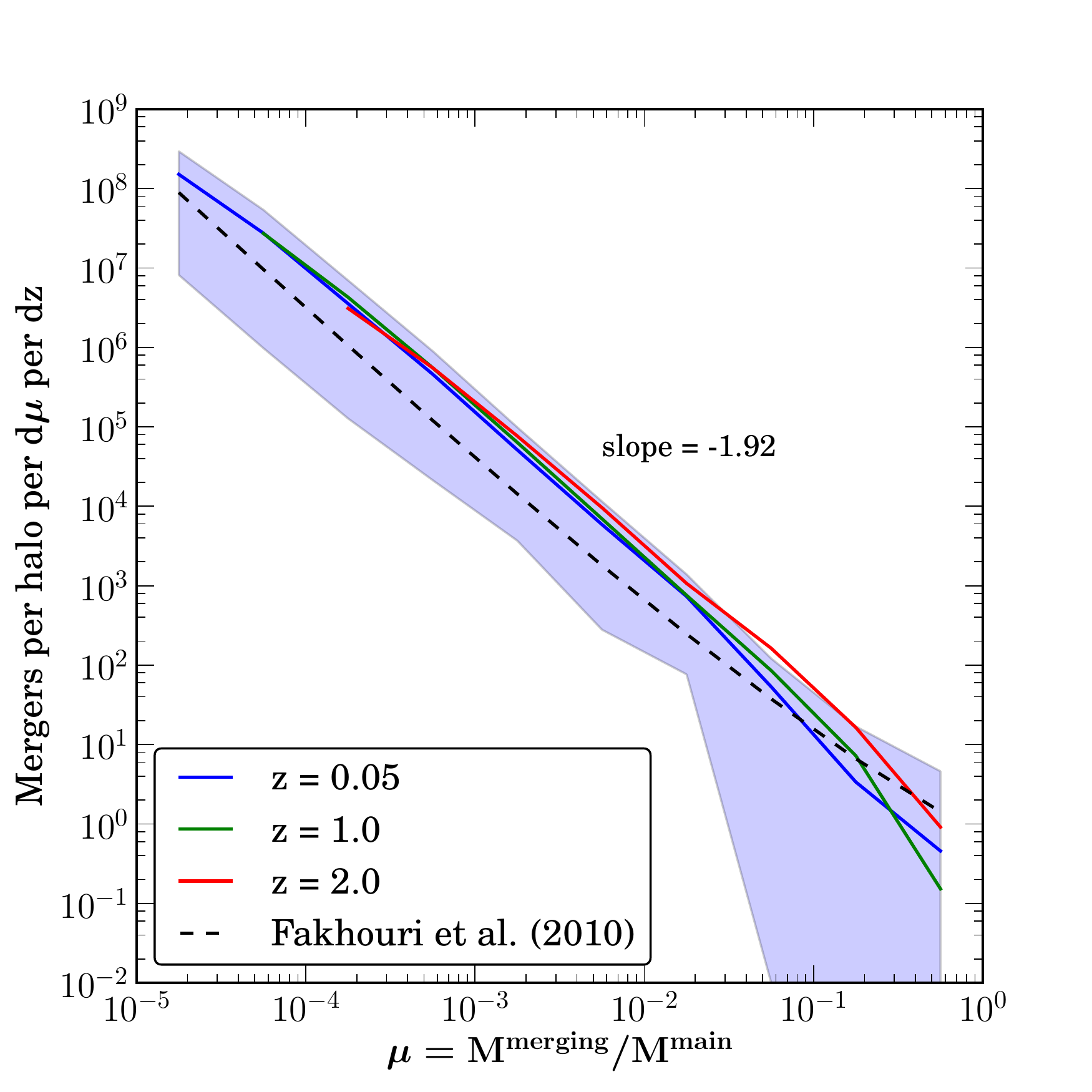}
\caption[]{Merger rate of {\sc Rhapsody} halos.  Left:
cumulative merger rate as a function of redshift (per halo).  The
average number of merger events each main halo has experienced since
a given $z$ is shown for three different merger mass ratios $\mu =
M^{\rm merging}/M^{\rm main}$.  Right: differential merger rate as a
function of merger mass ratio.  The number of merger
events per halo per $d\mu$ per $dz$ is shown as a function of $\mu$. The three
different curves correspond to different redshifts, and the trend is
independent of redshift.  For both panels, each colored band
corresponds to the standard deviation of the curve of the same color.}
\label{fig:MergerRate}
\end{figure*}

The merger rate experienced by a dark matter halo as a function of
time and merger ratio is a specific prediction of the $\Lambda$CDM
model \citep[e.g.,][]{PressSchechter74,LaceyCole93} and has been
calibrated with ever-improving precision using the extended
Press-Schechter model \citep[e.g.,][]{NeisteinDekel08} and $N$-body
simulations \citep[e.g.,][]{FakhouriMa08,Stewart09,Genel09}.  In this
work, we use the new merger tree implementation of
\cite{Behroozi11tree} to re-examine the merger rate of cluster-size
halos.  We consider the epoch of a merger event to be the earliest
redshift when the center of a smaller halo is inside the virial radius
of a larger halo, and the merger ratio, $\mu = M^{\rm merging}/M^{\rm
main}$, is defined at this epoch.

Figure~\ref{fig:MergerRate} presents the merger rate for {\sc
Rhapsody} halos.  The left panel shows the cumulative merger rate: The
$x$-axis specifies a look-back redshift, and the $y$-axis shows the
average number of merger events each main halo has experienced since
this redshift.  Different curves correspond to different merger mass
ratio thresholds, $\mu\geq$ 0.01, 0.03, 0.1, and 0.3, and the
shaded regions indicate the standard deviation of the sample.  We identify
the redshift of the last major merger, $z_{\rm lmm}$, using $\mu \geq
0.3$.  Our merger rate is consistent with the trend of the main halo
mass presented in \cite{Fakhouri10}, who have used Millennium
Simulations I and II and do not have sufficient statistics in our mass
regime.

The right panel shows the merger rate as a function of the merger mass
ratio $\mu$.  We plot the differential number of merger events each
main halo has experienced, per $d\mu$ per $dz$, for a given merger
ratio.  The different curves represent different redshifts.  We find
that the merger rate trends are almost invariant with redshift.  The blue-shaded
region corresponds to the scatter of the blue curve
($z=0.05$) and indicates the large variation of merger rate from halo
to halo.  The logarithmic slope, -1.92, is very close to the value -2
in \cite{FakhouriMa08} and \cite{Fakhouri10}.  We also plot the
fitting formula in \cite{Fakhouri10} as a black dashed curve
($z=0.05$). While the overall slope agrees well, their normalization
is slightly lower but still agrees within our error bar.

Figure~\ref{fig:cor} shows that $z_{\rm lmm}$ correlates with $\zhalf$, as well as various
halo structural properties, which will be detailed in the following sections.

\section{The Impact of formation time on the density profile}\label{sec:concentration}

The density profiles of dark matter halos have been shown to follow
the universal NFW form \citep{NFW97} and can
be well characterized by the concentration parameter $c$.  Calibrating
the concentration--mass relation and its scatter has been an ongoing
effort
\citep[e.g.,][]{Bullock01,Wechsler06,Neto07,Maccio08,Gao08,Prada11,Bhattacharya11}
and is of increasing importance for interpreting observations.  As
mentioned in the introduction, the CLASH project is a major effort of
the {\em Hubble Space Telescope} and aims for detailed and unbiased
measurements of the density profile of galaxy clusters, which are
tests of both the $\Lambda$CDM paradigm and our understanding of the
assembly of clusters.  In addition, the modeling of the
concentration--mass relation impacts the interpretation of the weak
lensing results \citep[e.g.,][]{King11} and X-ray results
\citep[e.g.,][]{Ettori10}.

In this section, we first provide fits to the density profiles of the
halos in the {\sc Rhapsody} sample. The halos of galaxy clusters,
being the most massive objects in the universe, assemble very late in
cosmic history, so that recent violent merger and accretion events are
expected to impact their density profiles. For this reason, we study
in detail the relation between their formation history and profiles in
the remainder of the section.

\subsection{Fitting the density profile}\label{sec:concentration_fit}
\begin{figure*}[t]
\centering
\includegraphics[width=\columnwidth]{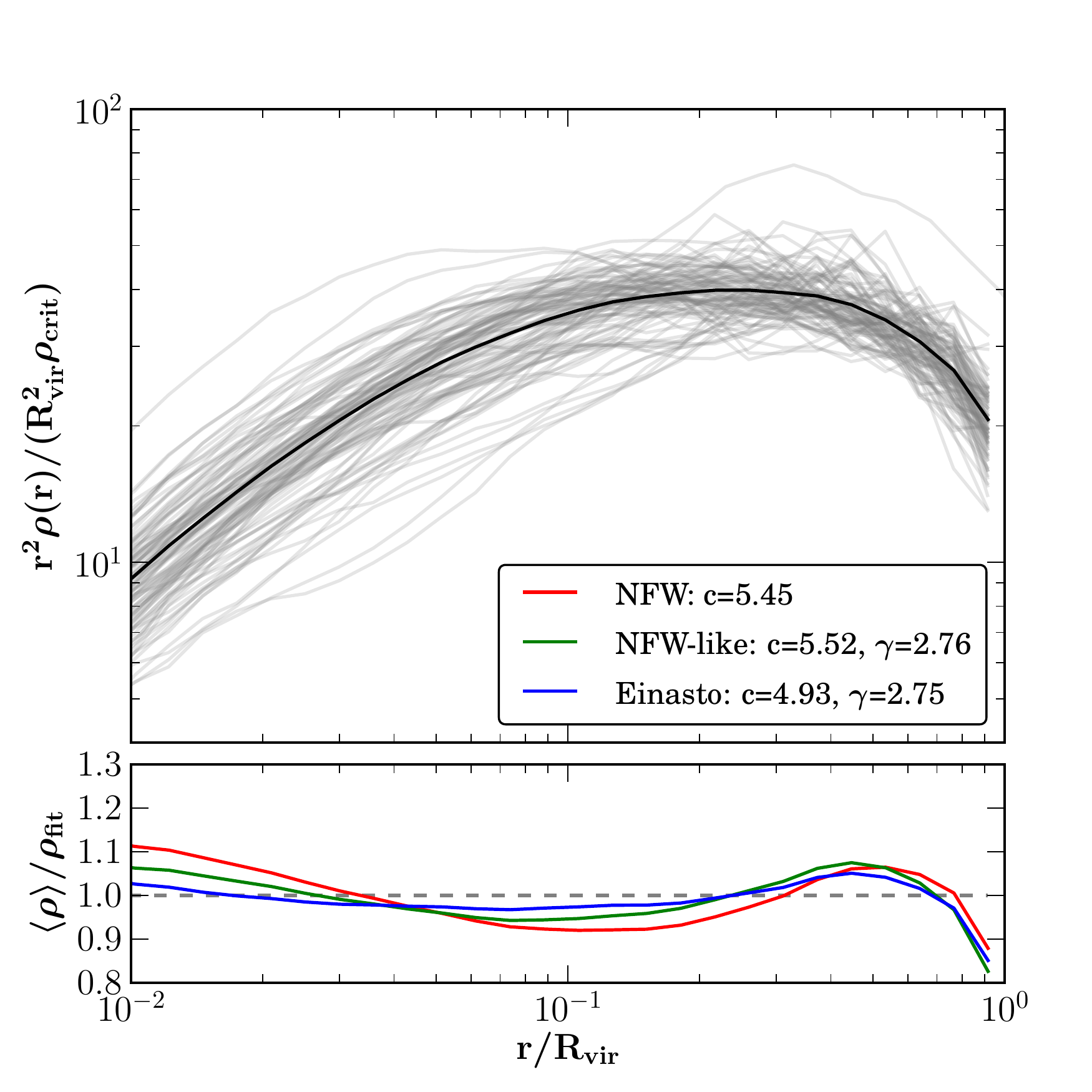}
\includegraphics[width=\columnwidth]{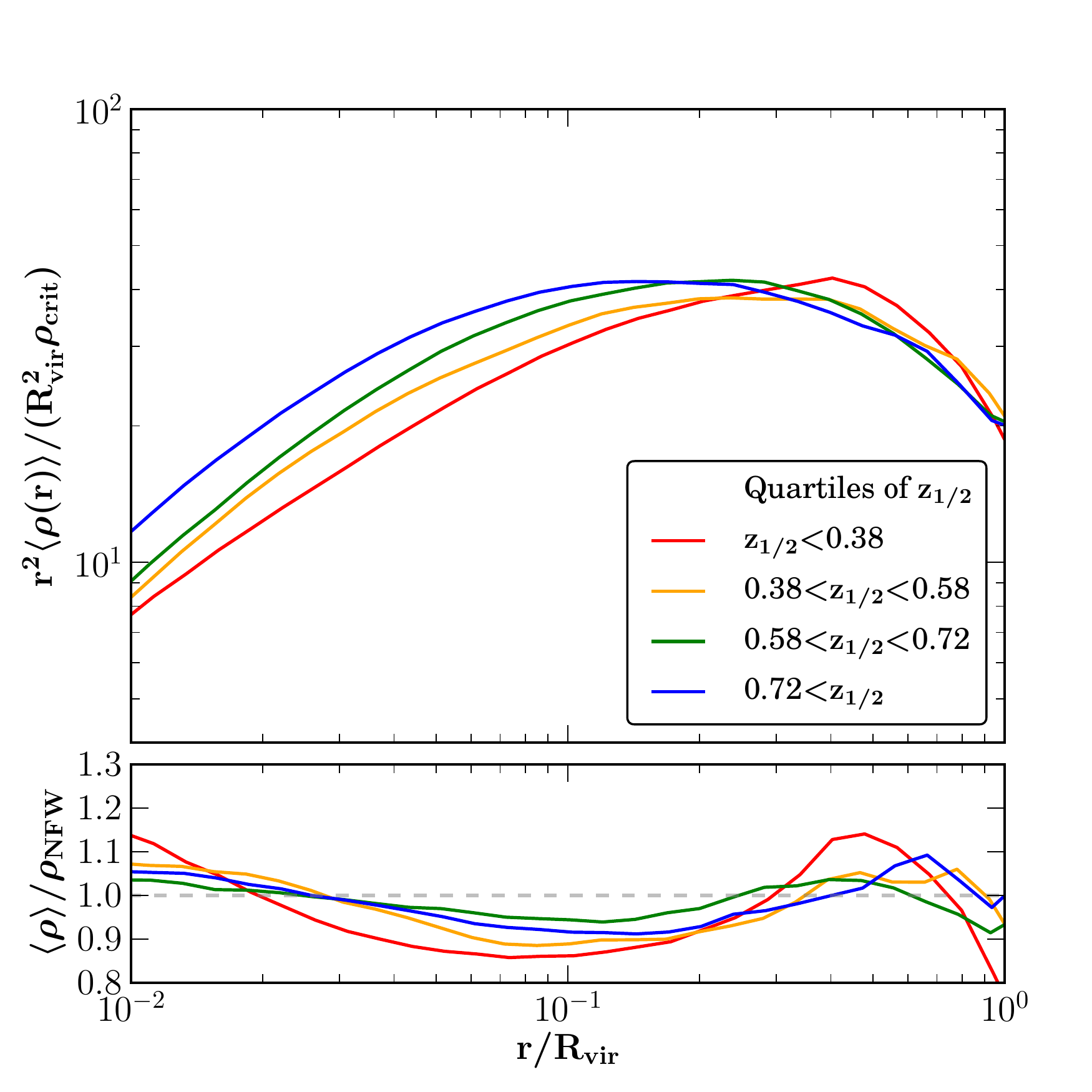}
\caption[]{Left: density profiles of the main halos in {\sc
    Rhapsody} (gray: individual halos; black: average).  The bottom
  panel shows the residuals with respect to three models (fits to the
  mean profile).  Right: the impact of formation time on density
  profile.  Halos are binned into quartiles based on their
  $\zhalf$, and the range of each quartile is indicated in the legend.
  The mean of halos in each quartile is shown in the
  upper panel.  The bottom panel shows the residuals with respect to
  the NFW profile.  Deviations from the NFW profile are systematic,
  with the largest deviation seen for late-forming halos and
  monotonous decrease when considering halos in the earlier forming
  bins.}
\label{fig:concentration}
\end{figure*}

The left panel of Figure~\ref{fig:concentration} shows the density
profiles of the main halos in {\sc Rhapsody}.  The black curve
corresponds to the mean density profile, while the gray curves
correspond to individual halos.  We use 32 bins between $10^{-2.5}
\Rvir$ and $\Rvir$ equally spaced in $\log r$ for plotting these
curves.  We note that the difference in the binned density profile
between each re-simulated halo and its low-resolution version is
between 5\% and 10\%.

For each main halo, we use all dark matter particles between 13
$\hikpc$ and $\Rvir$ to fit the density profile, adopting the
maximum-likelihood estimation without binning in radius (explained in
detail in Appendix \ref{app:concentration}).  This lower limit of
13 $\hikpc$ corresponds to 4 times the softening length, which is
slightly larger than the empirical scale below which the density
profile is not well resolved \citep[see, e.g.,][]{Ghigna00}.

There has been an ongoing effort to quantify the deviation from NFW
and to search for alternative fitting functions
\citep[e.g.,][]{Navarro04,Merritt06,Gentile07,Gao08,Navarro10}.  In
this work, we fit three parameterizations:
\begin{enumerate}
\item The NFW profile 
\beqa
\frac{\rho(r)}{\rho_{\rm crit}} &=& \frac{\delta_c}{(r/r_s)(1+r/r_s)^2} \
, \\
\frac{d\ln \rho}{d\ln r} &=& -\frac{1+3(r/r_s)}{1+(r/r_s)} \ ,
\eeqa
which is characterized by one parameter $c_{\rm NFW} = \Rvir/r_s$.
\item  An ``NFW-like'' profile with a free outer slope $\gamma$
\beqa
\frac{\rho(r)}{\rho_{\rm crit}} &=&
\frac{\delta_c}{(r/r_s)(1+r/r_s)^{\gamma-1}} \ ,\\
\frac{d\ln \rho}{d\ln r} &=& -\frac{1+\gamma(r/r_s)}{1+(r/r_s)} \ ,
\eeqa
which reduces to the NFW profile when $\gamma = 3$.  This profile can
be characterized by two parameters, $\gamma_{\rm NFW-like}$ and
$c_{\rm NFW-like} = (\Rvir/r_s) (\gamma_{\rm NFW-like}-2)$.  The
latter is defined so that $\Rvir/c_{\rm NFW-like}$ equals the radius
at which the density slope is $-2$.  We impose $r_s < \Rvir$ in the
fitting procedure to avoid possible divergence of $\gamma$.

\item The Einasto profile \citep{Einasto65}
\beq
\frac{d\ln \rho}{d\ln r} = -2 \left(
  \frac{r}{r_{-2}}\right)^{\alpha_{\rm Einasto}} \ .
\eeq
This model is characterized by two parameters, $r_{-2}$ and $\alpha_{\rm Einasto}$.  To
compare with the other two models, we define
\beqa
c_{\rm Einasto} &=& \frac{\Rvir}{r_{-2}} \ ,\\
\gamma_{\rm Einasto} &=& 2 c_{\rm Einasto}^{\alpha_{\rm Einasto}}\
\mbox{(slope at $\Rvir$)} \ ,
\eeqa
where $\gamma_{\rm Einasto}$ is the slope of the logarithmic density profile
at $\Rvir$ and can be compared with
$\gamma_{\rm NFW-like}$. 
\end{enumerate}

The best fit values of these parameters are summarized in
Table~\ref{tab:properties}, and the comparison of the goodness-of-fit 
is detailed in Appendix \ref{app:concentration}.
For the NFW fit, our mean value agrees with
\cite{Prada11} and \cite{Bhattacharya11}.  Our
standard deviations are $\sigma({c})/c = 0.26$ and $\sigma(\log_{10}
c)=0.11$, which are slightly smaller than the values quoted in
\cite{Bhattacharya11} (0.33 and 0.16 based on $\Delta_{200c}$).  This
difference is presumably due the decreasing scatter in the
concentration--mass relation with increasing mass; as mentioned in
\cite{Bhattacharya11}, their scatter is slightly smaller for massive
halos (for $M_{200} > 8\times10^{14}\ \hiMsun$, the scatter is
$\sigma(c)/c = 0.28$, which is very close to our value).

In the left panel of Figure~\ref{fig:concentration}, we add a bottom
panel to compare the residual of these three models,
$\avg{\rho}/\rho_{\rm fit}$.  Here $\rho_{\rm fit}(r)$ is obtained by
fitting the {\em stacked} binned density profile of all halos, and the
legend shows the best-fit parameters for each model.  We note that
these values are slightly different from the average values of halos
(shown in Table~\ref{tab:properties}); i.e., the fit of the average
results in slightly higher concentrations than the average of the fit
to each halo.  Among these three profiles, the Einasto profile fits to
the stacked density profile best, deviating by up to 5\%, whereas the
NFW profile deviates by up to 10\%.  A similar trend of deviations has
also been shown in the Phoenix simulations \citep{Gao12} and the
Aquarius simulations \citep{Navarro10}.

\subsection{Density profile and formation history}\label{sec:c_zhalf}
\begin{figure*}
\centering
\includegraphics[width=\columnwidth]{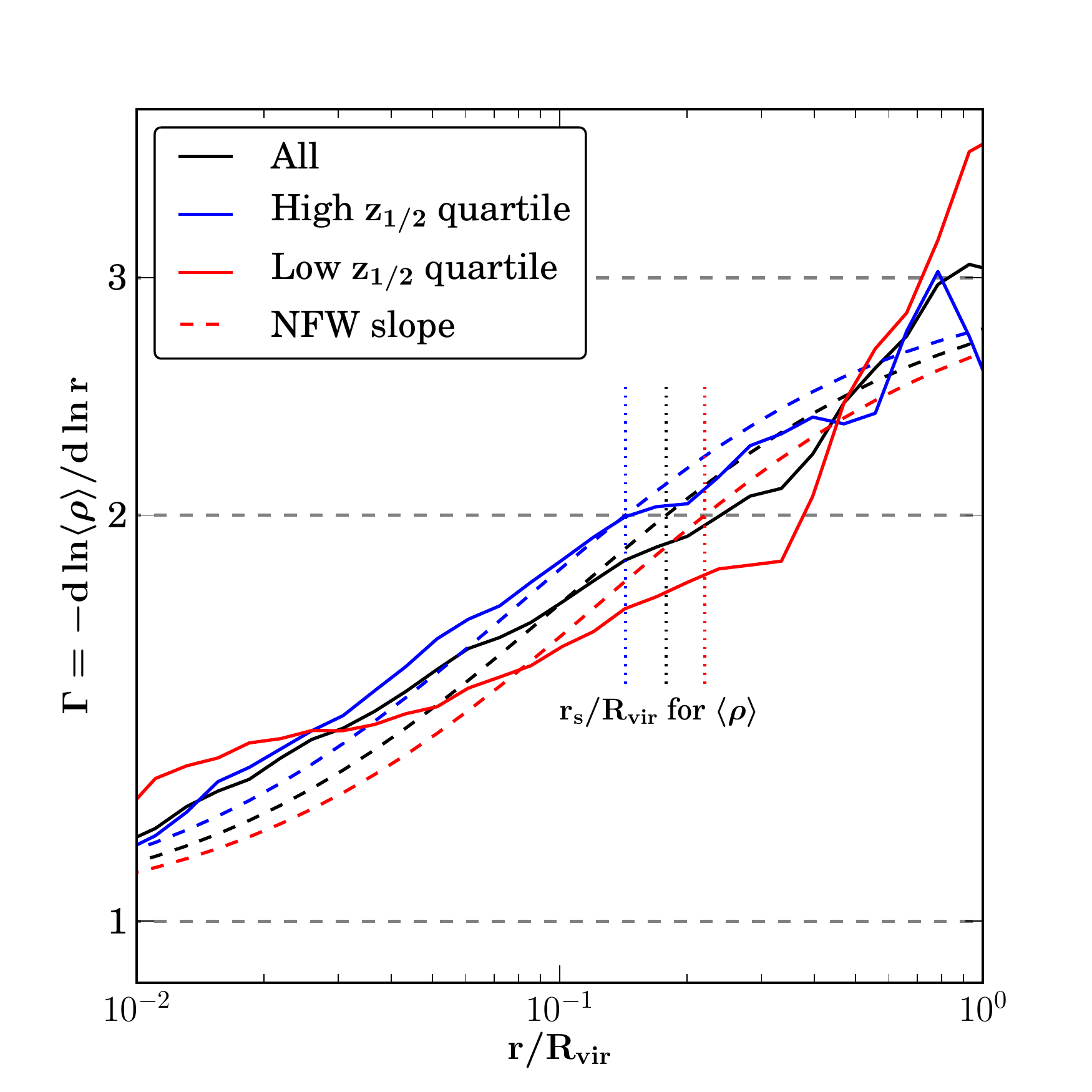}
\includegraphics[width=\columnwidth]{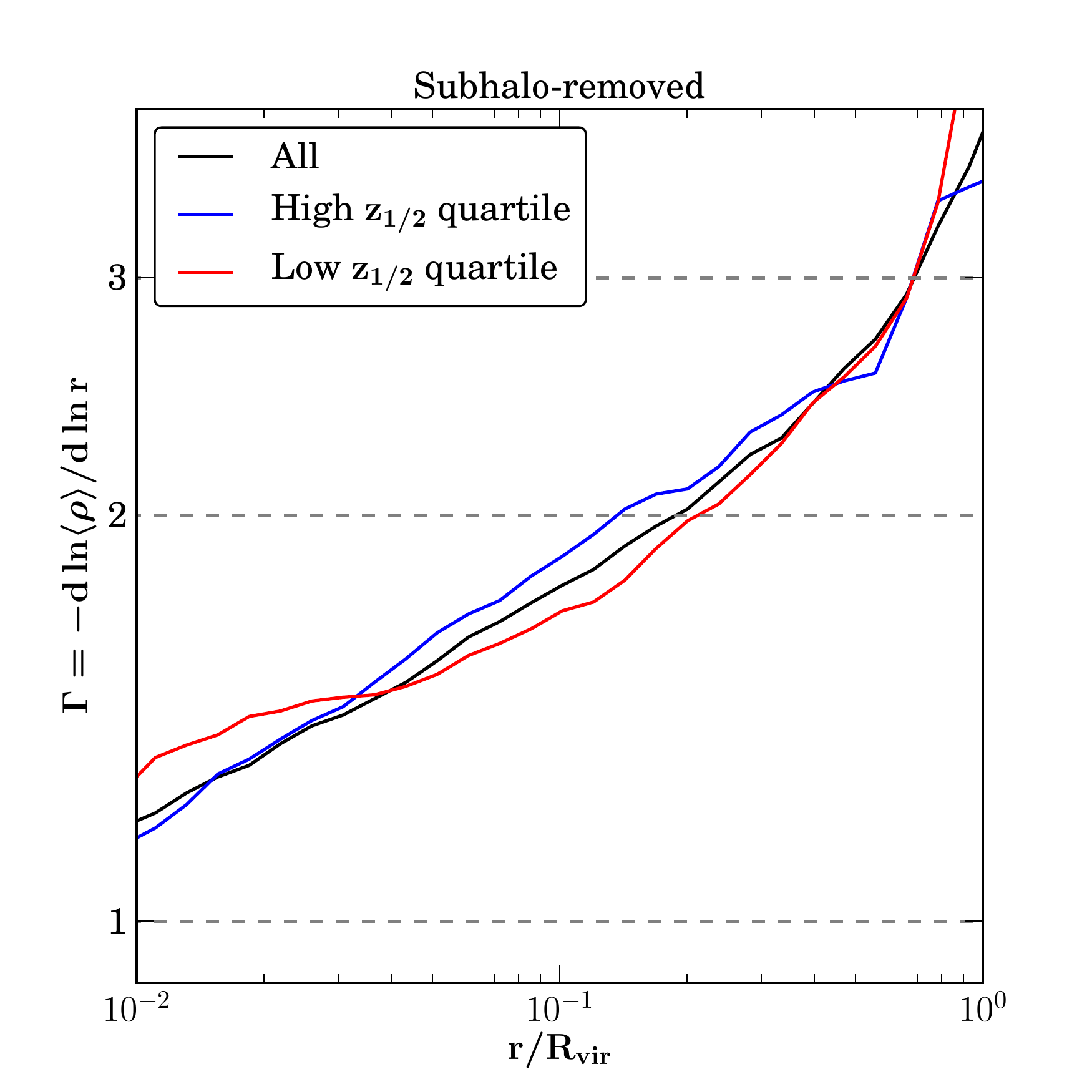}
\caption[]{Left: the absolute value of the logarithmic slope of
the density profile, $\Gamma$.  The red/blue curve corresponds to the
mean value of the low/high $\zhalf$ quartile.  We also mark the $r_s$
and slope expected from NFW.  Except for the early-forming halos at
around $r_s$, $\Gamma$ deviates substantially from NFW.  For
late-forming halos, $\Gamma < 2$ at $r_s$ (shallower than NFW
expectation).  For early-forming halos, $\Gamma$ approximately follows
a power law and can be well described by an Einasto profile.  For
late-forming halos, $\Gamma$ appears to be a broken power law and
suddenly increases around 0.3 $\Rvir$.  Right: halos with their
massive subhalos ($\vmax > 200\ \kms$) removed.  For late-forming
halos, the kink around 0.3$\Rvir$ disappears with the removal of
subhalos, and their differences from the early-forming halos are
reduced.  This result indicates that the differences in $\Gamma$ in
the left panel are largely driven by massive subhalos.  }
\label{fig:slope}
\end{figure*}
\begin{figure*}
\centering
\includegraphics[width=\columnwidth]{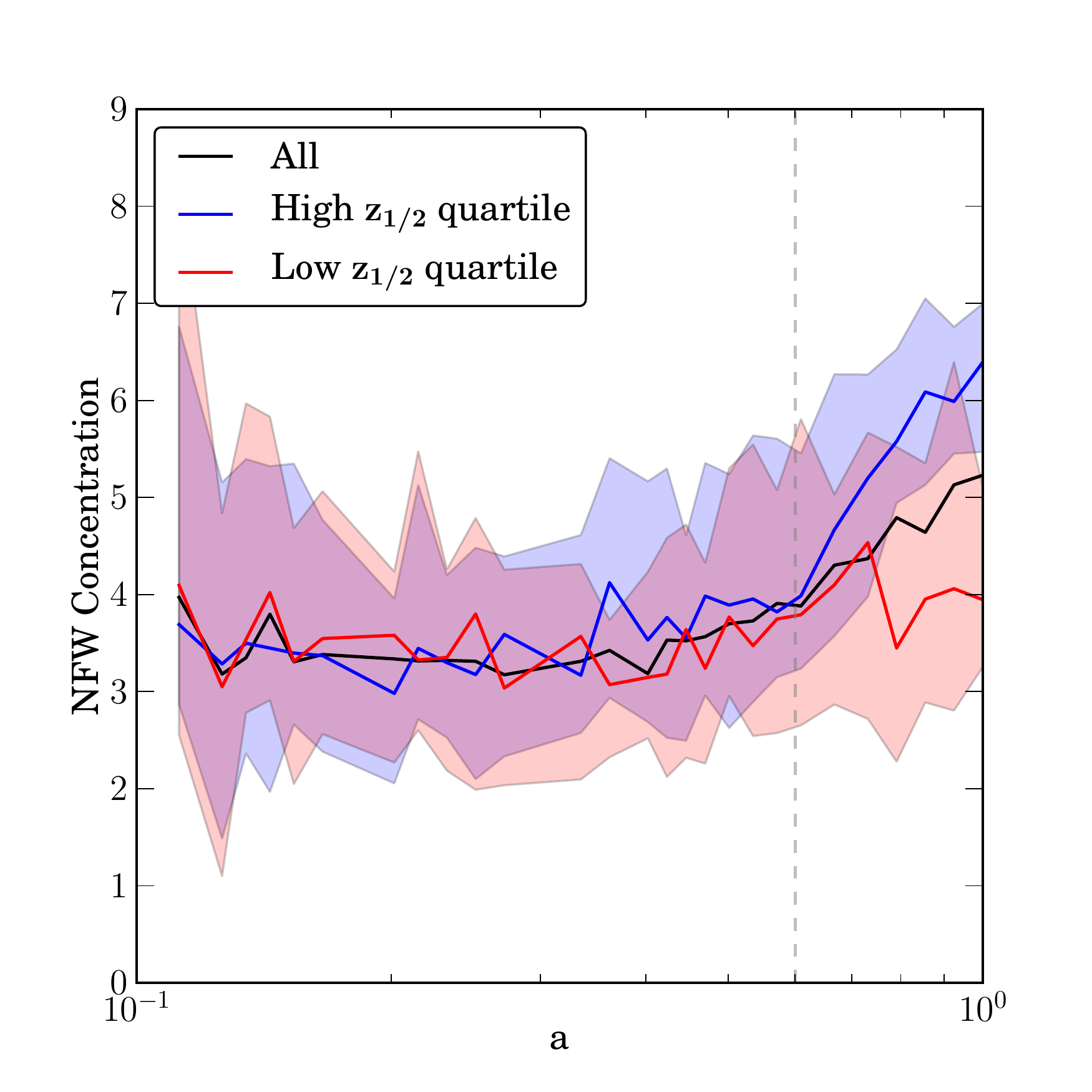}
\includegraphics[width=\columnwidth]{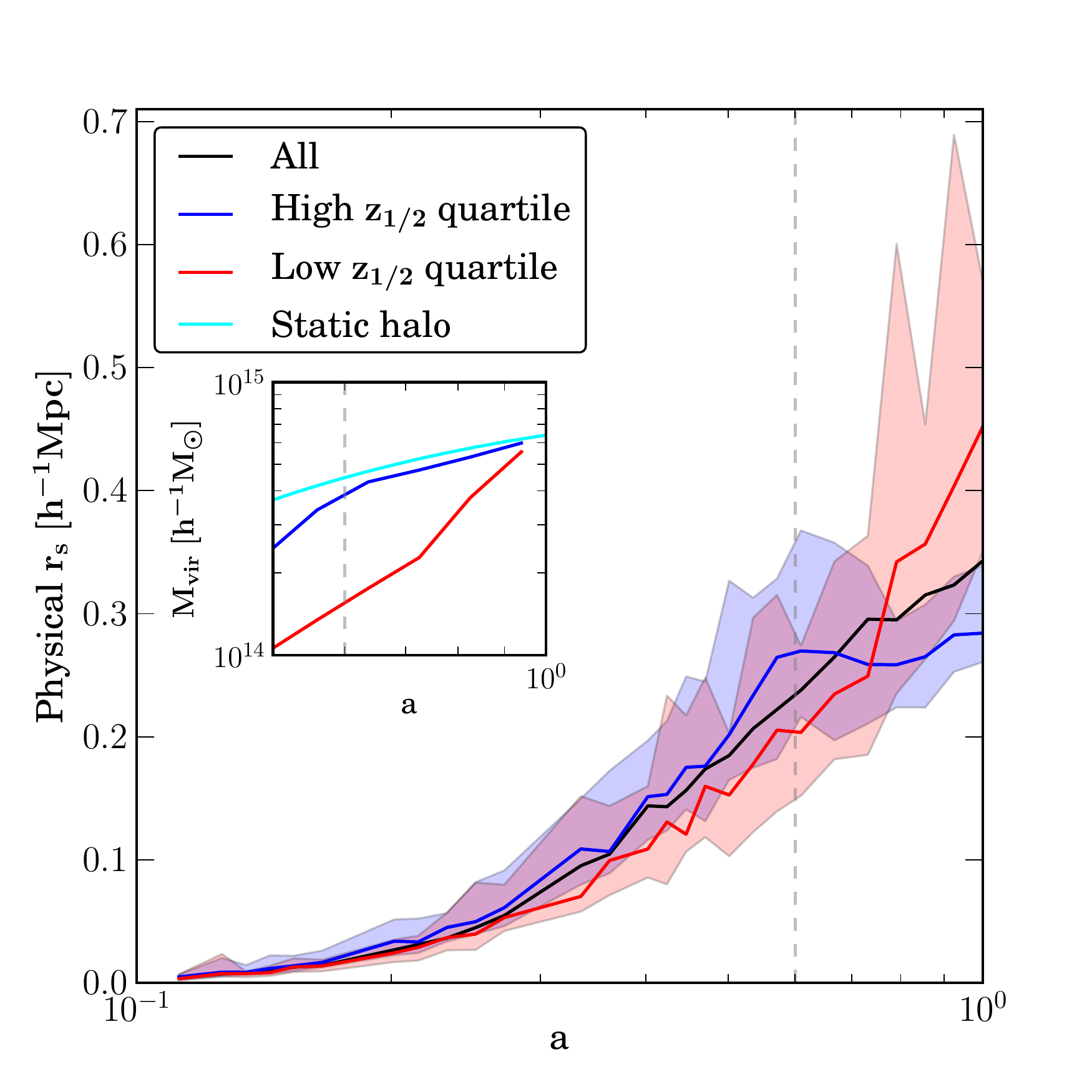}
\caption[]{Evolution of NFW concentration (left) and scale
radius (right).  The red/blue curve corresponds to the mean of
the halos in low/high $\zhalf$ quartile.  For late-forming halos, the
mean concentration remains relatively constant for all redshifts.  For
early-forming halos, the mean concentration remains constant at high
redshift; however, for $a>0.6$, the concentration steadily increases,
while $r_s$ remains nearly constant. In the inset, we compare their
late-time mass accretion with the pseudo-evolution of a static halo,
indicating that the increase in concentration of early-forming halos
is largely driven by pseudo-evolution.
}
\label{fig:c_z}
\end{figure*}

We explore several aspects of the impact of the formation history on
halo density profiles.  In Section \ref{sec:c_cor}, we present the
correlation between concentration and formation time, as well as the
impact of formation time on the deviation from NFW.  In
Section \ref{sec:slope}, we show that the slope of profile also depends on
formation time and that subhalos alter the slope profile of
late-forming halos.  We show the evolution of concentration in
Section \ref{sec:c_z}.

\subsubsection{Formation time, concentration, and the deviation from NFW}\label{sec:c_cor}

Figure~\ref{fig:cor} includes $c_{\rm NFW}$ and the goodness-of-fit of
NFW, defined as
\beq
\Delta_{\rm NFW} = {{\rm max} | M(<r) -M_{\rm NFW}(<r)|}/{\Mvir} \ .
\eeq
We find that both quantities are strongly correlated with $\zhalf$,
$z_{\rm lmm}$, and $\gamma-\beta$, and the correlation is strongest with
$\zhalf$.  The correlation between concentration and formation time is
well known \citep[e.g.,][]{NFW97,Wechsler02}, and the canonical
explanation is that the concentration of halos reflects the physical
density of the universe at their formation epoch.  The correlation
between the deviation from NFW and formation time can be understood
through the relaxedness of halos: late-forming halos tend to show
larger deviation from NFW because of recent mergers and late-time mass
accretion.  Although highly concentrated clusters tend to be closer to
NFW, we note that $c_{\rm NFW}$ and $\Delta_{\rm NFW}$ only have a
weak correlation.

To further explore the correlation between formation time and
concentration, as well as the deviation from NFW, we split our halos
into quartiles based on $\zhalf$.  In the right panel of
Figure~\ref{fig:concentration}, we present the average density profile
of halos of each quartile.  The upper panel shows that the formation
time can lead to systematic differences in the density profile.  For
each average density profile, we fit the NFW profile and present the
residual, $\avg{\rho(r)}/\rho_{\rm NFW}$. in the bottom panel.
Deviations from NFW are again systematic; around 0.1 $\Rvir$ NFW fits
tend to overestimate.  The latest-forming quartile tends to have the
largest deviations from NFW, while the two early-forming quartiles
show less deviation.  We note that the deviation from NFW cannot be
entirely attributed to the departure from spherical symmetry.  As we
will show in Section \ref{sec:shape}, the shape and triaxiality of halos do
not strongly correlate with formation time and cannot account for the
observed deviation from NFW.

\subsubsection{The slope of density profile: Impact of subhalos}\label{sec:slope}

To further understand the impact of formation time on the deviation
from NFW, we compare the logarithmic slope of the density profile in
the quartiles of highest and lowest $\zhalf$ ($\zhalf>0.72$
  and $\zhalf < 0.38$, respectively).  The left panel of
Figure~\ref{fig:slope} shows the absolute value of the logarithmic
slope of the density profile, $\Gamma = -d\ln\avg{\rho}/d\ln r$, where
$\avg{\rho}$ is obtained by stacking halos in the highest and lowest
$\zhalf$ quartiles.  We also show the slope expected from NFW as
dashed curves.  The slopes deviate substantially from NFW for most
radii.

The effect of formation time on the halo concentration can also be
seen in the left panel of Figure~\ref{fig:slope}.  We mark the
location of $r_s$ as vertical dotted lines.  For early-forming halos,
$r_s$ is exactly the radius where the slope equals $-2$.  In contrast,
for late-forming halos, $r_s$ is smaller than the radius where the
slope equals $-2$.  This indicates that the NFW model does not provide
an adequate fit to their profiles, leading to the correlation between
$\zhalf$ and $\Delta_{\rm NFW}$ seen in Figure~\ref{fig:cor}.  In
addition, we note that a lower $\Gamma$ around 0.2$\Rvir$ corresponds
to a larger $r_s$, and thus a smaller NFW concentration.

In the notation of the Einasto profile,
\beqa
&&\Gamma =2\left( \frac{r}{r_{-2}} \right)^{\alpha_{\rm Einasto}} \ ,\\
&&\frac{d\ln\Gamma}{d\ln r} = \alpha_{\rm Einasto} \ .
\eeqa
For early-forming halos, $\Gamma$ is close to a power law; therefore,
the Einasto profile provides a fit better than NFW.\footnote{We
calculate the residual for Figure~\ref{fig:slope}:
$\Delta^{\Gamma}_{\rm model} = \sum_{i=1}^{N} {|\Gamma_{\rm true}(r_i) -
\Gamma_{\rm model}(r_i)|}/{\avg{\Gamma_{\rm true}}}$.  For
early-forming halos, $\Delta^{\Gamma}_{\rm NFW} = 1.2$ and
$\Delta^{\Gamma}_{\rm Einasto} = 0.65$.  For late-forming halos,
$\Delta^{\Gamma}_{\rm NFW} = 3.0$ and $\Delta^{\Gamma}_{\rm Einasto} =
2.9$.  For all halos, $\Delta^{\Gamma}_{\rm NFW} = 1.59$ and
$\Delta^{\Gamma}_{\rm Einasto} = 1.02$.  }  However, for late-forming
halos, neither model provides a good fit, and the logarithmic slope of
$\Gamma$ has a sudden increase around 0.3 $\Rvir$.  This ``kink'' in
$\Gamma$ can be explained by the presence of subhalos.  In the right
panel of Figure~\ref{fig:slope}, we present the slope with the massive
subhalos ($\vmax>200\ \kms$) removed.  After removing these subhalos,
the kink in the red curve disappears, and the difference between the
blue and red curves is significantly reduced.\footnote{Our subhalo
removal procedure is based on the subhalo-particle assignment based on
{\sc Rockstar}; different halo finders tend to have different
subhalo-particle assignment criteria \citep[e.g., see][]{Onions12}.
Since our purpose is to understand the trend with and without the
presence of the subhalos, we do not expect that detailed particle
assignment will impact the trend discussed here.}  We also note that
in the presence of a massive subhalo, the slope can be shallower near
the locus of the subhalo and steeper at larger radii.  Therefore, the
presence of subhalos can explain the kink in the slope in
Figure~\ref{fig:slope}.

\subsubsection{Evolution of concentration}\label{sec:c_z}

Figure~\ref{fig:c_z} shows the mean concentration evolution of our
main halos.  The red/blue curve corresponds to the low/high $\zhalf$
quartile.  To emphasize the late-time phenomenon, we plot logarithmic
scale factor as the $x$-axis.  For $a<0.6$, the averages of both
populations remain relatively constant and have similar values,
reflecting the split in $\zhalf$.  For $a>0.6$, the concentration of
early-forming halos increases steadily with time and the scatter
decreases; this steady increase in concentration is presumably related
to the lack of significant mass growth and the resulting higher degree
of relaxedness.  On the other hand, the mean concentration of
late-forming halos does not increase much and still shows large
scatter.

The right panel of Figure~\ref{fig:c_z} shows the evolution of $r_s$
in physical $\hiMpc$.  When the mean concentration of early-forming
halos increases, the mean $r_s$ value remains nearly constant,
suggesting that the increase in concentration could be mainly
driven by the pseudo-evolution of $\Rvir$ related to the time
dependence of $\Deltavir\rho_{\rm crit}$.  To confirm this, we add an
inset showing the mean mass evolution of both populations as well as
the pseudo-evolution of a static halo.  As can be seen, for
early-forming halos, the mass evolution for $a>0.6$ is close to the
pseudo-evolution, indicating that these halos are on average close to
static and their increase in concentration is indeed driven by the
pseudo-evolution.  We note that the constant $r_s$ has been shown in
lower mass systems \citep[e.g.,][]{Bullock01,Wechsler02}, here we show
that it also happens in relaxed massive halos.

\section{The impact of formation time on the phase-space structure}\label{sec:Vr}

In this section, we present the impact of formation time on the
phase-space structure of halos.  The phase-space structure has been
explored for the purpose of seeking alternative definitions of halo
boundary and mass \citep[e.g.,][]{Busha05,Cuesta08}; understanding the
connection between density and velocity distribution
\citep[e.g.,][]{HansenMoore06,Navarro10}; calculating possible
caustics to aid the search for lensing or dark matter annihilation
signal \citep[e.g.,][]{Diemand08b}.  In this work, we focus on the
radial velocity profile because the infall and outflow patterns carry
the information of formation and merger history and can help us
understand the build-up of the density profile as well as the state of
equilibrium.

\begin{figure}[t]
\centering
\includegraphics[width=\columnwidth]{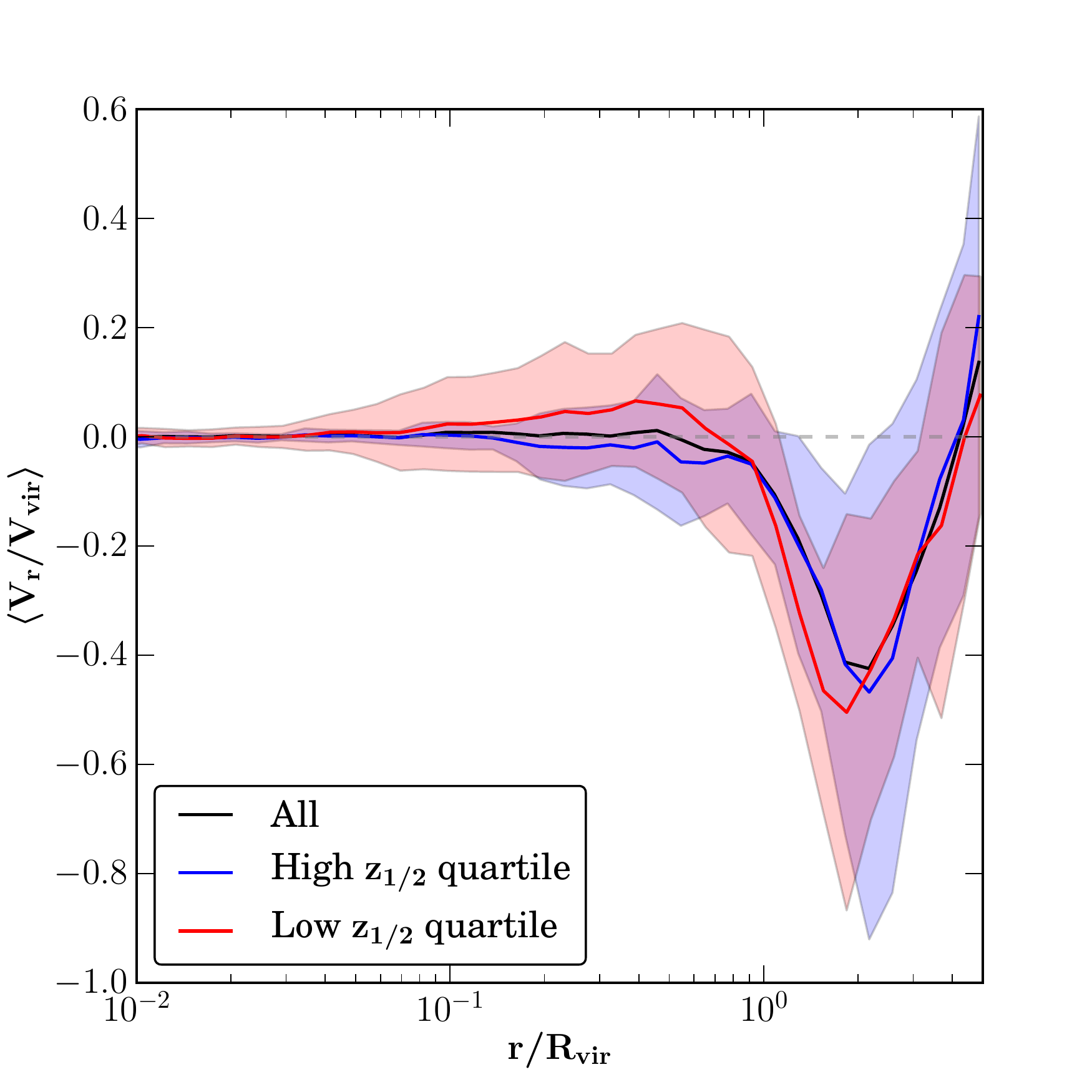}
\caption[]{Mean radial velocity profiles of {\sc Rhapsody} halos.
The red/blue curve corresponds to halos in the low/high $\zhalf$
quartile, and the band corresponds to the standard deviation of the
sample.  Late-forming halos show a significant outflow between 0.3 and
1 $\Rvir$ on average and have larger scatter.  In addition,
late-forming halos tend to have stronger infall than early-forming
halos beyond $\Rvir$.  We will demonstrate that the outflow of the
late-forming halos is related to the coherent motions of dark matter
particles associated with subhalos.}
\label{fig:Vr}
\end{figure}
\begin{figure*}[t]
\centering
\includegraphics[width=\columnwidth]{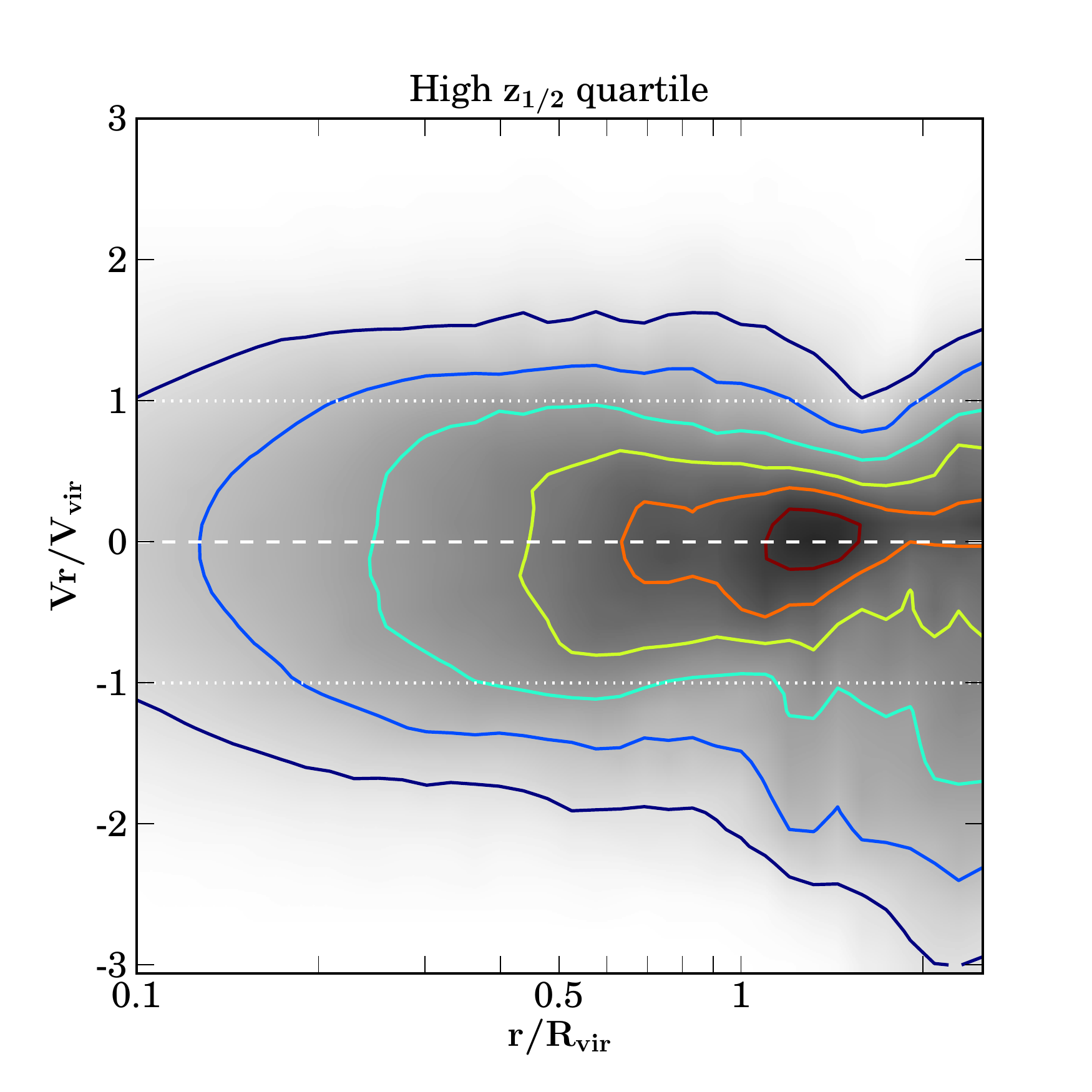}
\includegraphics[width=\columnwidth]{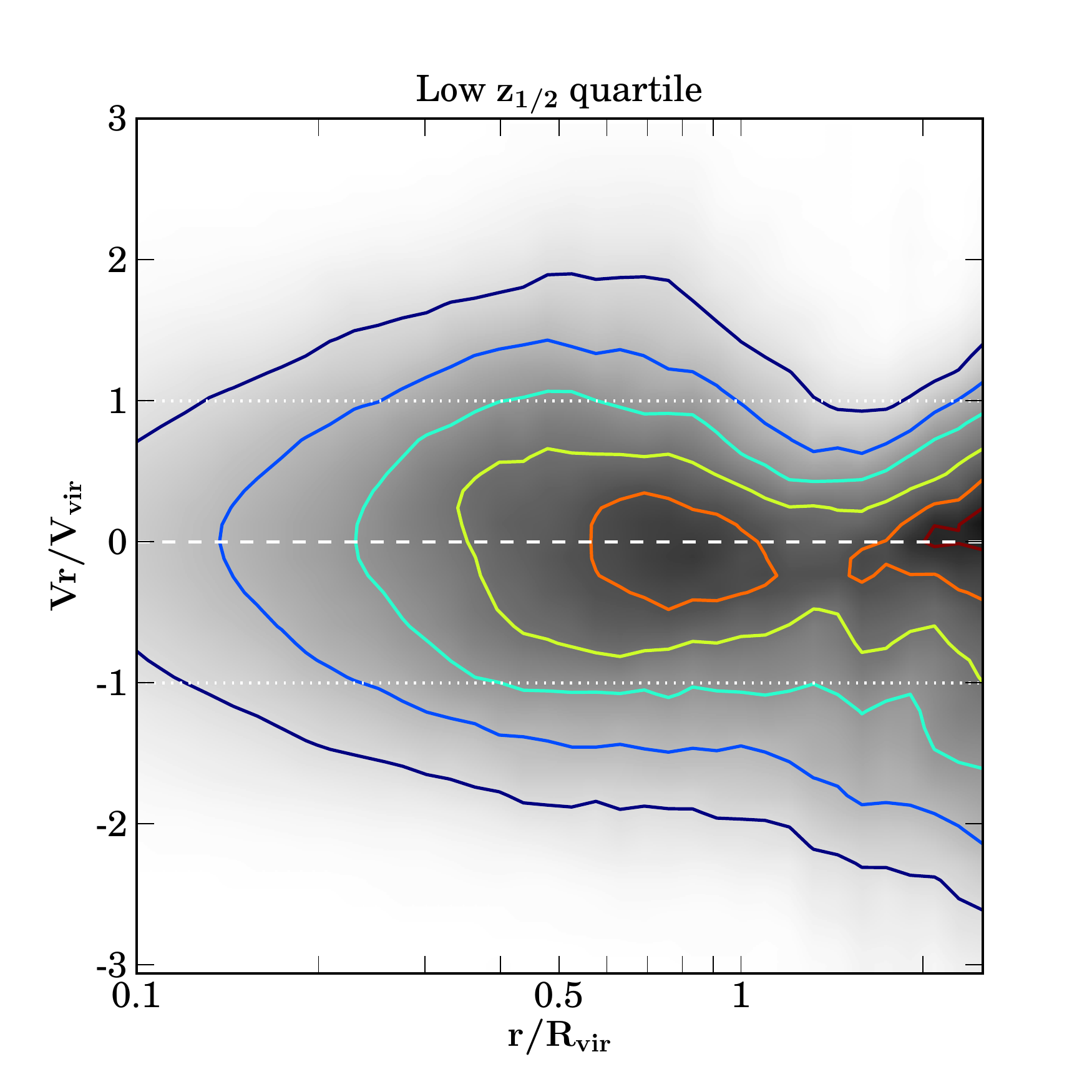}\\
\includegraphics[width=\columnwidth]{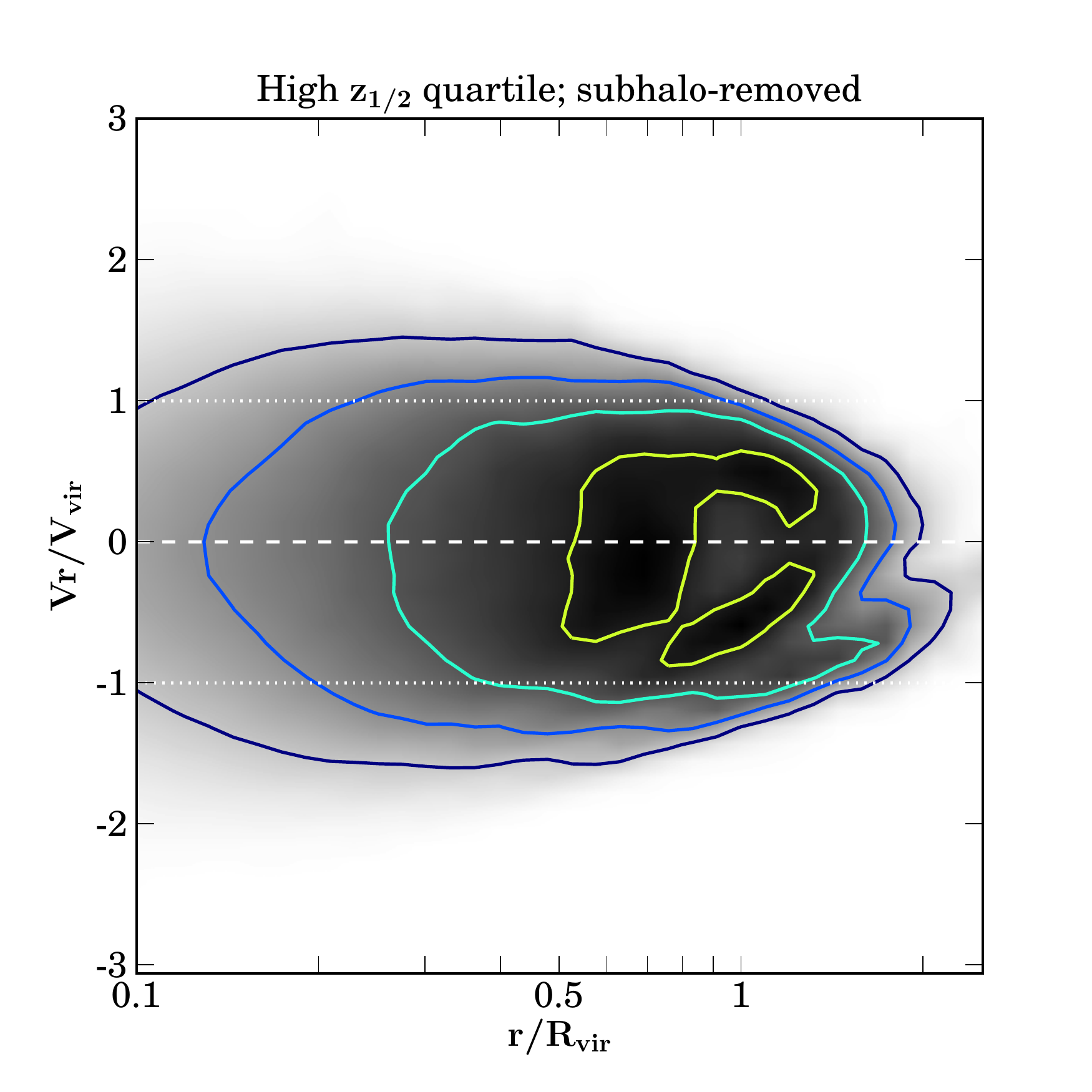}
\includegraphics[width=\columnwidth]{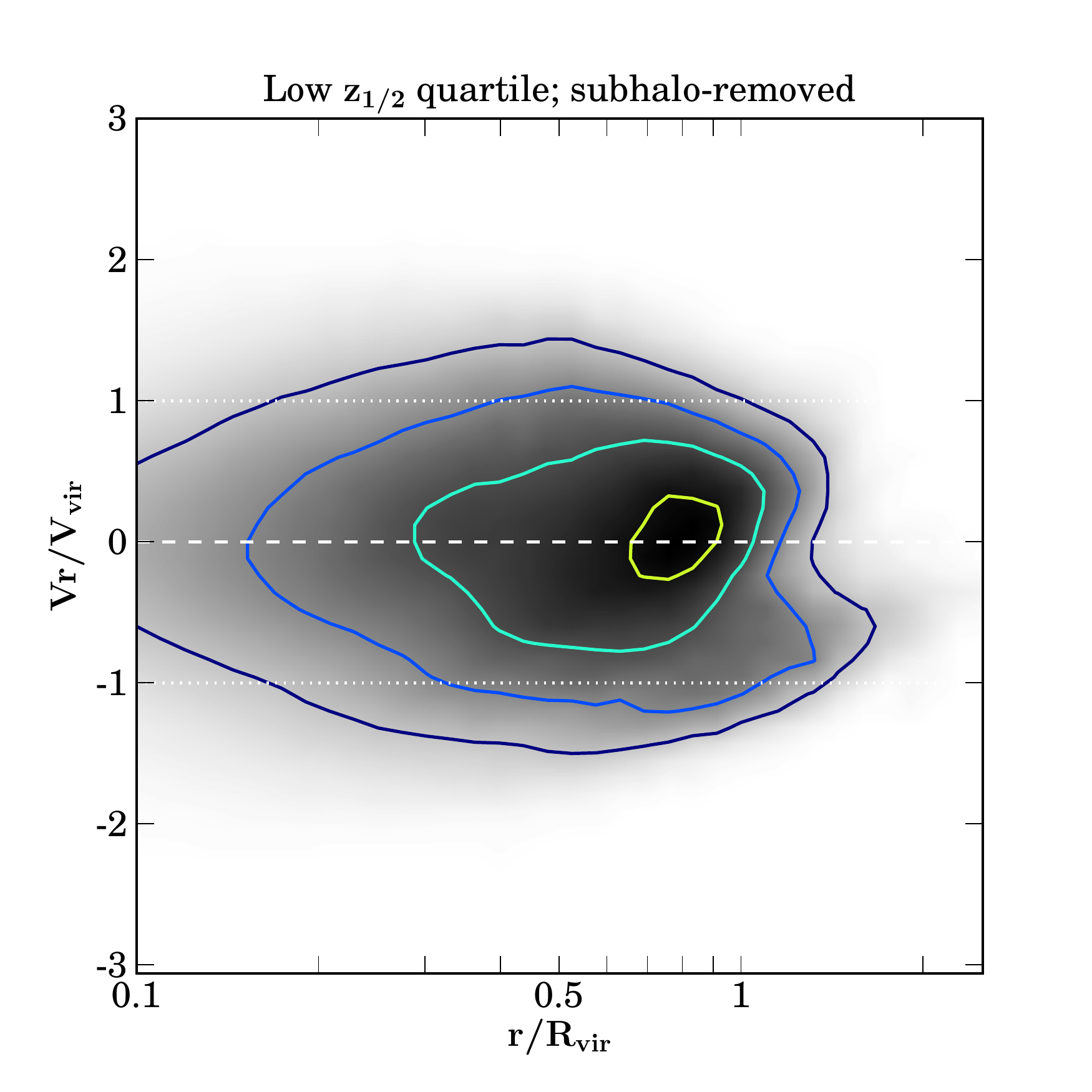}
\caption[]{Impact of formation time on the phase-space structure
of halos.  Upper/lower: including/excluding subhalos.  
Left/right: the high/low $\zhalf$ quartile.  Comparing the two upper
panels, we can see that the late-forming halos show stronger outflow
between 0.3 and 1 $\Rvir$, and the curvy shapes of the contours in
this region correspond to the kink in the slope of density profile at
the same scale shown in Figure~\ref{fig:slope}.  Comparing the two
right panels, we can see that the outflow in the subhalo-removed
systems is reduced.}
\label{fig:Vr_r_contour}
\end{figure*}

Figure~\ref{fig:Vr} presents the differential radial velocity profile 
\beq
\left\langle \frac{V_r}{V_{\rm vir}} \left( \frac{r}{\Rvir} \right) \right\rangle
\eeq
of the dark matter particles that are bound to the main halos.  We
again split halos by their $\zhalf$ and present the halos in the
highest and lowest quartiles.\footnote{We note that for each
halo, $\avg{V_r}$ is itself the mean of the particles in a radial bin,
and the stacking process is the average over the $\avg{V_r}$ of
individual halos.} The late-forming halos (red curve) show an average
outflow within $\Rvir$.  As will be discussed in the following
paragraph, these outflows are related to the coherent motions of the
particles associated with recently merged subhalos.  When these
subhalos pass though the main halo, their particles tend to have high
outflowing velocities.  On the other hand, early-forming halos tend to
show more regular infall patterns.

The presence of a significant fraction of particles exceeding the
virial velocity of the main halo is consistent with the detection of
the so-called ``back-splash'' population of halos
\citep[e.g.,][]{Gill05,Ludlow09} in the outskirts of massive
halos. These halos have left the virial radius after their initial
accretion onto the halo and have experienced tidal stripping much like
their siblings remained inside the virial radius.

Figure~\ref{fig:Vr_r_contour} compares the phase-pace structures for
early-forming and late-forming halos.  We stack the halos in the high
(left panel) and low (right panel) $\zhalf$ quartiles and present
$V_{r}$ versus \ $r$ (both normalized using the virial units).  These two
populations show different features in the phase-space diagram.  The
late-forming halos again show an average outflow between 0.3 and 1
$\Rvir$.  In addition, in this radial range, the shapes of the
contours are different.  For early-forming halos, the contours are
close to parallel to the $x$-axis; for late-forming halos, the
contours are curved upward.

We find that the outflow in late-forming halos, like the kink in
$\Gamma$, is related to the presence of massive subhalos.  In the
bottom panels of Figure~\ref{fig:Vr_r_contour}, we remove subhalos
with $\vmax > 200 \ \kms$, and the outflow is reduced.

Although formation history does not strongly impact the velocity
ellipsoid (as will be shown in Section \ref{sec:VelEllip}), it clearly
impacts the radial velocity profile.  We also note that the outflows
shown here are more likely to be found in dynamically young systems,
including cluster-size systems with late formation time (like our
sample) or galactic systems at high redshift.

\section{Shapes and alignments}\label{sec:shape}

In this section, we present the shapes for the spatial distributions
and velocities of dark matter particles of the main halos in {\sc
Rhapsody}, as well as the alignments of their orientations.  Our
motivation is again related to the formation history: The accretion of
matter onto massive halos is presumably correlated with the
surrounding large-scale structure, and anisotropic accretion onto the
halo can leave an imprint in both the shape of the halo and the motion
of matter in its interior.

\subsection{Halo shapes}\label{sec:PosShape}

$N$-body simulations have shown that dark matter halos are generally not
spherical objects but have significant ellipticity and triaxiality
\citep[e.g.,][]{JingSuto02,KasunEvrard05,Allgood06,Bett07,Hayashi07}.
Calibrations of halo shapes from simulations directly impact the
accuracy of weak lensing mass calibration
\citep[e.g.,][]{Corless07,Bett12, Schneider12} and can be used to
constrain the self-interaction cross-section of dark matter particles
\cite[e.g.,][]{Miralda-Escude02}.

The shape parameters are typically defined through the mass
distribution tensor with respect to the halo center
\beq
I_{ij} = \langle r_i r_j\rangle \ ,
\label{eq:shape}
\eeq
where $r_i$ is the $i$th component of the position vector ${\bf r}$ of a
dark matter particle with respect to the halo center, and the average
$\langle \cdot \rangle$ is over all dark matter particles within
$\Rvir$.  Since all particles within $\Rvir$ are of the same mass, no
weighting by mass is needed.  The eigenvalues of $I_{ij}$ are sorted
as $\lambda_1 > \lambda_2 > \lambda_3$, and the shape parameters are
defined as $a=\sqrt{\lambda_1}$, $b=\sqrt{\lambda_2}$,
$c=\sqrt{\lambda_3}$.  We present the dimensionless ratios $b/a$ and $c/a$.
In addition, the triaxiality parameter is defined as
\beq
T = \frac{a^2-b^2}{a^2-c^2} \ .
\label{eq:triax}
\eeq

We list these halo shape properties in Table~\ref{tab:properties} and
present the distribution of $c/a$ in Figure~\ref{fig:cor}.  The shape
parameter $c/a$ is only weakly correlated with formation history.  The
correlation is even weaker for $b/a$ and $T$.

It has been shown that the halo shape depends on the radius at which
it is measured \citep[e.g.,][]{BailinSteinmetz05}.  We also measure
the halo shape at $R_{500c}$ and find that $b/a$, $c/a$, and $T$
measured at $R_{500c}$ are correlated with those measured at $\Rvir$,
with correlation coefficients 0.61, 0.63, and 0.60, respectively.  The
major axes measured with $R_{500c}$ and $\Rvir$ have a median angle of
21$^\circ$, which agrees with the value ($\sim 20^\circ$) reported by
\cite{Schneider12} for the Millennium Simulations.

\cite{Allgood06} used a smaller radius $0.3\Rvir$ and an iterative
method to calculate the reduced inertia tensor (weighted by
$r^{-2}$ to suppress the influence of the larger radii). They
found that $c/a$ is correlated with formation time, and the
correlation is weaker for higher mass.  In our measurement
with unreduced inertia tensor at $\Rvir$ and
$R_{500c}$, the correlation between $c/a$ and $\zhalf$ is not strong (0.27 and 0.29 respectively).  Since
\cite{Allgood06} did not state the correlation coefficient and did not
have statistics in our mass regime, we cannot make a direct comparison
but note that the different mass regime and measurement methods could
impact these correlations. \cite{Bett12} recently showed that the
shape measured from the iterative reduced tensor is similar to and
only slightly more spherical than those obtained with the simple
method applied here.

\cite{Shaw06} showed that the measured shapes depend somewhat on the
state of relaxedness (also see \citealt{SkibbaMaccio11,WongTayler12}). 
For our sample of halos, we also find a weak
correlation between $c/a$ and the goodness-of-fit proxy $\Delta_{\rm
NFW}$. While the correlation is weak, it is however interesting to
note that none of our halos is close to spherical (high $c/a$) and has
a high $\Delta_{\rm NFW}$ at the same time.  Given such a correlation,
one might thus wonder whether the deviations from the NFW
profile---with a strong dependence on the formation time $\zhalf$, as
discussed in Section \ref{sec:c_z}---are driven by systematic
deviations from sphericity of the entire virialized halo, or by the
anisotropic distribution of the most massive subhalos. To decide this,
we repeat the measurements of the shape parameters after removing all
subhalos with $\vmax > 200\ \kms$. We find that the correlation
coefficient between $\Delta_{\rm NFW}$ and $c/a$ at $\Rvir$ drops from
$-0.20$ to a mere $0.01$ when subhalos are removed. Similarly, at
$R_{500c}$, we observe a drop of the correlation coefficient from
$-0.29$ to $0.05$. This is strong evidence that the deviations from
NFW at variance with formation time are predominantly driven by
recently accreted subhalos. The correlation between shape and
formation time then is a simple consequence of the anisotropic
distribution of these subhalos.

\subsection{Velocity ellipsoid}\label{sec:VelEllip}

\cite{WhiteM10} have demonstrated that the anisotropic motion of
subhalos in clusters introduces significant scatter in the velocity
dispersions measured along different lines of sight.  Here we follow
the same procedure to measure the properties of the velocity ellipsoid
of the dark matter particles in the {\sc Rhapsody} sample.

Analogous to the mass distribution tensor, the velocity ellipsoid is
defined as
\beq
\sigma^2_{ij} = \langle v_i v_j\rangle \ .
\label{eq:VelEllip}
\eeq
Sorting the eigenvalues of the velocity ellipsoid as $\lambda_1 >
\lambda_2 > \lambda_3$, one can again define shape parameters $a^{(v)}
= \sqrt{\lambda_1}$, $b^{(v)} = \sqrt{\lambda_2}$, $c^{(v)} =
\sqrt{\lambda_3}$, and dimensionless ratios $b^{(v)}/a^{(v)}$ and
$c^{(v)}/a^{(v)}$ to describe the velocity ellipsoid.

The mean and scatter of the velocity dispersions measured along
different lines of sight can be calculated as
\beqa
\langle \sigma^2_{\rm los} \rangle &=&
\frac{1}{3}(\lambda_1+\lambda_2+\lambda_3) \ ,\\
(\delta \sigma^2_{\rm los})^2 &=& \frac{4}{45} (\lambda_1^2+\lambda_2^2+\lambda_3^2-\lambda_1\lambda_2-\lambda_2\lambda_3-\lambda_3\lambda_1) \ . \ \ \ \ \ \ \
\label{eq:sig_los}
\eeqa
We list these parameters in Table \ref{tab:properties}.  In
Figure~\ref{fig:cor}, we show the distribution of $c^{(v)}/a^{(v)}$,
which, like $c/a$, is only weakly correlated with formation time.  The
correlation is similar for $b^{(v)}/a^{(v)}$ and $\delta
\sigma^2_{\rm los}$.

We also measure the velocity ellipsoid based on $R_{500c}$. In this
case, the correlation is stronger than halo shape: $b^{(v)}/a^{(v)}$,
$c^{(v)}/a^{(v)}$, and $\delta \sigma^2_{\rm los}$ measured at $R_{500c}$
are correlated with those values measured at $\Rvir$ with correlation
coefficients 0.80, 0.86, and 0.86, respectively.  The major axes of
the velocity ellipsoid measured with $\Rvir$ and $R_{500c}$ have a
median angle of 14$^\circ$.

As expected, the ellipsoid of dark matter distribution and velocity
ellipsoid are correlated.  While $b/a$ and $b^{(v)}/a^{(v)}$ have a
correlation of 0.35, we find a stronger correlation between $c/a$ and
$c^{(v)}/a^{(v)}$ of magnitude 0.53. We note that the velocity
ellipsoid is in general more spherical than velocity ellipsoid, this
difference is related to the fact that the shape of the gravitational
potential is in general more spherical than density distribution.
This trend was discussed in \cite{KasunEvrard05}, who used the Hubble
volume simulation and defined halos based on $\Delta_{200c}$.
Although our simulation uses a different mass definition and is in a
different mass regime from theirs, our axis ratios for spatial
distribution and velocity are in agreement within 0.01 with theirs.
In addition, the major axes of the spatial distribution and velocity
ellipsoids have a median angle of 27$^\circ$, which is slightly larger
than the value 22$^\circ$ in \cite{KasunEvrard05}.

We do not find significant correlation between any of these shape
parameters with the environmental parameters we measured (including
several halo number overdensity and nearest neighbors for several mass
thresholds).  In Paper II, we will repeat these measurements for the
spatial distributions and velocities of subhalos and compare them with
the results measured from all dark matter particles shown here.

\section{Summary and Discussion}\label{sec:summary}

We have presented the first results of the {\sc Rhapsody} project,
which includes 96 cluster-size halos with mass $\Mvir = 10^{14.8\pm
0.05}\hiMsun$, re-simulated from 1 $h^{-3}\rm Gpc^3$ with a resolution
equivalent to 8192$^3$ particles in this volume.  In addition to
achieving high resolution and large statistics simultaneously, {\sc
Rhapsody} is unique in its well-resolved subhalos and wide span of
evolution history.  {\sc Rhapsody} also implemented the
state-of-the-art algorithms for initial conditions ({\sc Music}), halo
finding ({\sc Rockstar}), and merger trees.

Our findings are summarized as follows.

\begin{enumerate}

\item {\em Properties of the cluster halos.} We have summarized the
  key properties (including various mass definitions, formation
  history proxies, halo concentration, and shape parameters) of the 96
  main halos of the {\sc Rhapsody} sample in
  Table~\ref{tab:properties} and shown the distributions and mutual
  covariance of a subset of these properties in Figure~\ref{fig:cor}.

\item {\em Mass accretion history and merger rate.} We have
  investigated the mass accretion history of the main halos in
  Section \ref{sec:MAH}, tracking the most massive progenitors over 5
  decades of mass growth.  We have shown that an exponential form does
  not provide an adequate fit to such a wide span of time, and an
  extra power law is needed.  In Section \ref{sec:merger}, we have shown
  that the differential merger rate follows a power law scaling with
  the merger mass ratio and is independent of redshift, in agreement
  with earlier work based on less massive halos or smaller samples.

\item {\em Density profile.} In Section \ref{sec:concentration}, we have quantified
  in detail how formation time impacts the halo density profile.  We
  have shown that the deviations from the NFW model systematically
  depend on the formation history of the halo.  Specifically,
  late-forming halos tend to show larger deviations from NFW.  In
  Section \ref{sec:slope}, we investigate the slope of the density profile.
  Early-forming halos can be well described by the Einasto model.  For
  late-forming halos, the slope appears to be a broken power law with
  a sudden transition around 0.3 $\Rvir$.  This kink in the slope
  profile is related to the abundant massive subhalos in these
  late-forming halos.  We have also shown in Section \ref{sec:c_z} that, for
  early-forming halos, the evolution of concentration between $z \sim
  1$ and 0 is consistent with pseudo-evolution.

\item {\em Halo shape and alignment.} As discussed in
 Section \ref{sec:shape}, the shape and velocity ellipsoid of our sample is
 only moderately correlated with the formation history.  We present the
 correlation and alignment of shapes and velocity ellipsoids measured
 at $\Rvir$ and $R_{500c}$.  Our results indicate that anisotropy of
 the entire virialized halo alone cannot explain the deviations of the
 density profiles from the NFW model and its dependence on formation
 time.

\item{\em Phase-space structure and formation time.} 
  We have provided evidence in Section \ref{sec:Vr} for a connection between
  the density profile, massive subhalos, and the formation
  history of the halos. When investigating
  the $r$--$V_r$ phase space, we find that late-forming
  halos show evidence for large fluctuations in the radial
  velocities of their dark matter particles, with a considerable fraction
  in excess of the virial velocities, leading to an outflow of mass
  from these cluster halos. We associate fluctuations in
  the radial velocities, outflows, as well as the kink in the density
  profile slope with recently accreted subhalos. 
  On the contrary, early-forming halos show a
  substantially more regular motion of dark matter, as expected if
  they are dynamically more relaxed. Our results thus indicate that
  the recently accreted subhalos drive the observed deviation
  from the NFW profile.  Due to their anisotropic distributions, these subhalos
  also lead to a correlation between formation time and halo shape.

\end{enumerate}

In Paper II, we analyze in detail the properties of the subhalo
population of the cluster halos.  In addition to the impact of
formation time on subhalo properties, we investigate the intertwining
effects from subhalo selection, stripping, and resolution of
simulations.

The {\sc Rhapsody} simulation suite provides valuable information for
other aspects of cluster cosmology.  For example, the cluster-size
halos in {\sc Rhapsody} can be further used to study the covariances
between mass tracers, for example, galaxy content, dynamics of
galaxies, weak gravitational lensing, X-ray, and the SZ effect.  The
formation history and environment of clusters can potentially impact
these mass proxies systematically, either by altering the intrinsic
properties of clusters or by affecting the observed properties through
the line-of-sight projection. As current multi-wavelength surveys
combine these different observables for cluster mass calibration, it
is imperative to understand the covariances between these observables.

Our re-simulation technique can also be applied to study the ``pink
elephant'' clusters, which refer to a handful of massive clusters
recently discovered at high redshift
\citep[e.g.,][]{Jee09,Foley11}. These clusters have stimulated a great
amount of discussion about whether they pose a challenge to the
current $\Lambda$CDM paradigm of cosmology
\citep[][]{Mortonson11,Hoyle11}. To interpret the cosmological
implications of these clusters correctly, it is important to
understand their mass calibration.  An extension of the current {\sc
Rhapsody} sample that includes a statistical sample for these massive
clusters at high redshift will improve our understanding of these
massive clusters.  Understanding the covariances between different
observable quantities and the potential biases in the mass
measurements of these clusters can help us disentangle the
astrophysical and cosmological implications of these clusters.

\acknowledgements This work was supported by the U.S.\ Department of
Energy under contract numbers DE-AC02-76SF00515 and DE-FG02-95ER40899,
and by Stanford University through a Gabilan Stanford Graduate
Fellowship to H.W. and a Terman Fellowship to R.H.W.  Additional support
was provided by SLAC-LDRD-0030-12.  This work uses data from one of
the {\sc LasDamas} simulations; mock galaxy catalogs from these
simulations are available at http://lss.phy.vanderbilt.edu/lasdamas/.
We are grateful to our {\sc LasDamas} collaborators, and especially
Michael Busha, who ran the {\sc Carmen} box that was used for
re-simulation; we also thank Michael for extensive helpful
discussions.  We thank Ralf Kaehler for assistance with visualizations
of the simulations.  We gratefully acknowledge the support of Stuart
Marshall, Amedeo Perazzo, and the SLAC computational team, as well as
the computational resources at SLAC.  We also thank Gus Evrard,
Eduardo Rozo, Matt Becker, Andrey Kravtsov, Sarah Hansen, Anja von der Linden, Douglas
Applegate, and Will Dawson for helpful discussions and comments.

\appendix

\section{Fitting the formation history}\label{app:MAH}
\begin{figure*}[t]
\includegraphics[width=0.33\columnwidth]{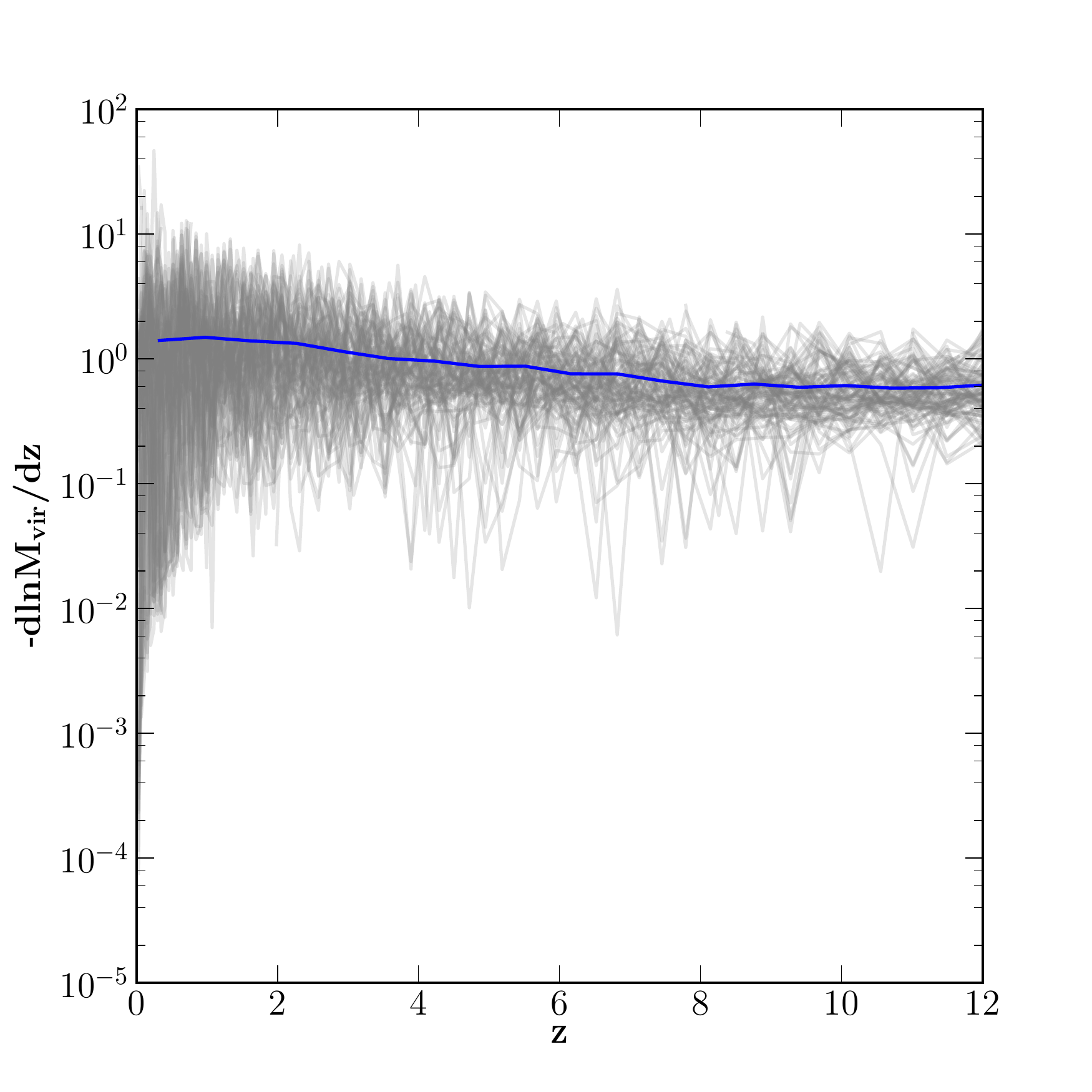}
\includegraphics[width=0.33\columnwidth]{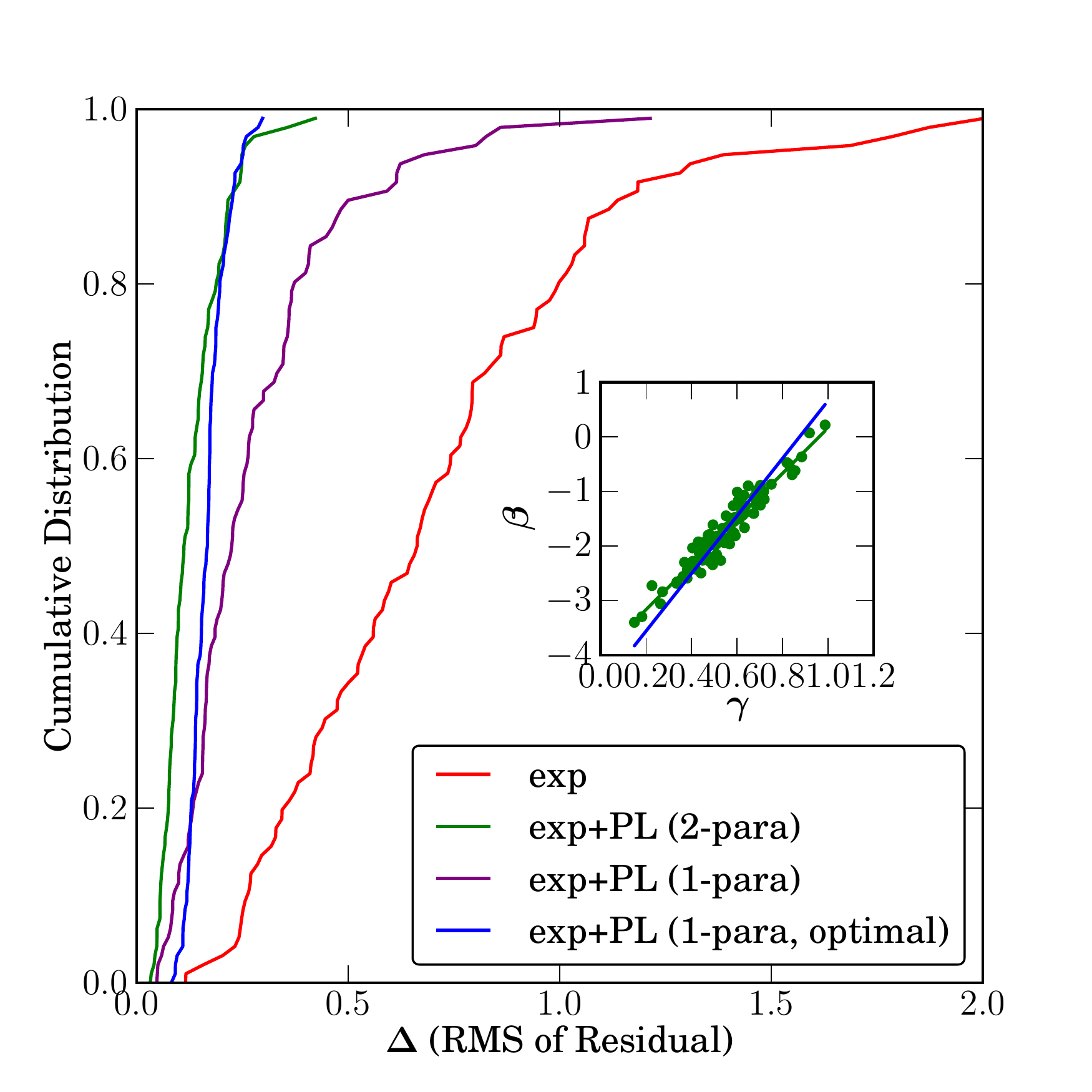}
\includegraphics[width=0.33\columnwidth]{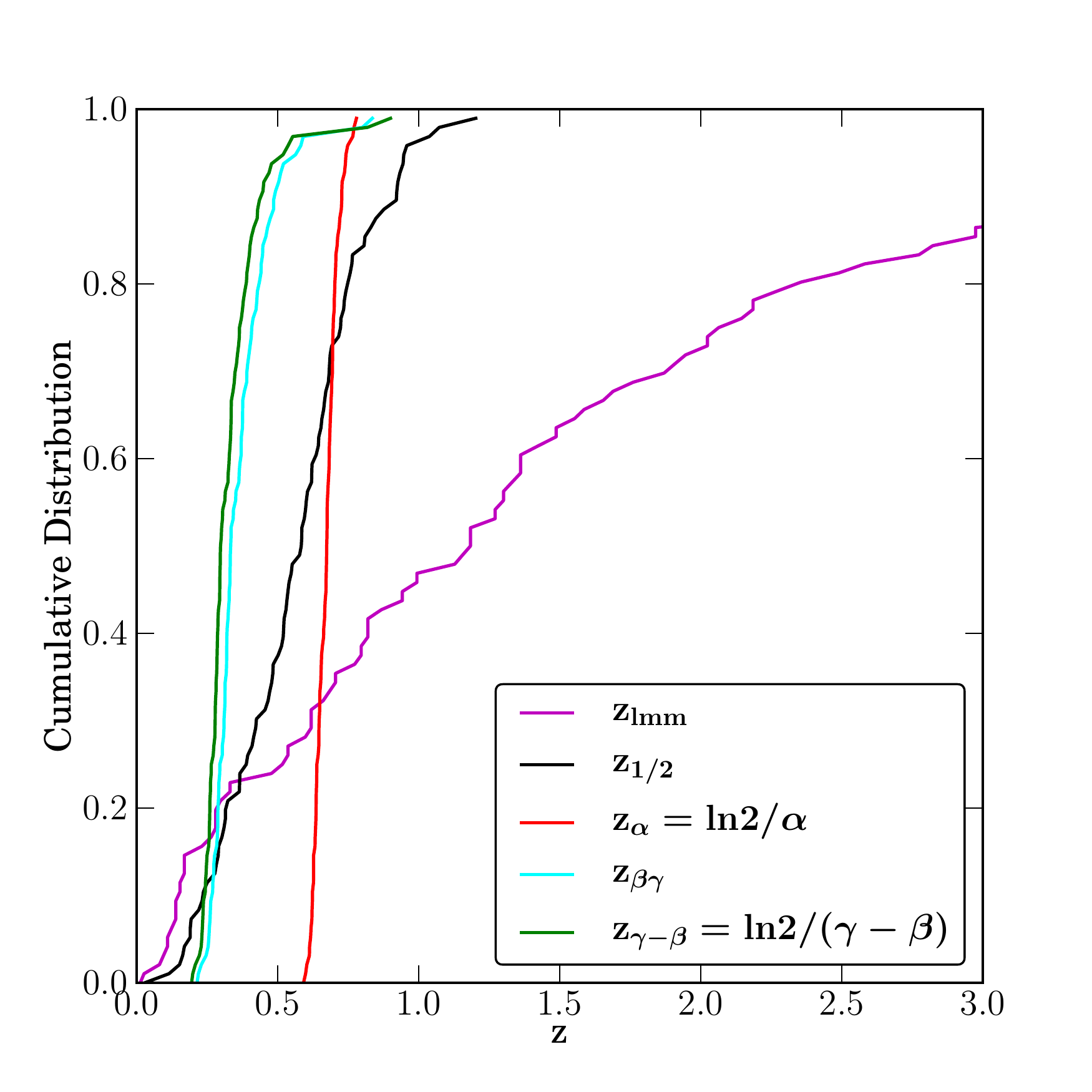}
\caption[]{
  Left: the mass accretion rate of the {\sc Rhapsody} halos, $-d\ln
  \Mvir/dz$, as a function of $z$, averaged over every 3 output redshifts to reduce the noise.
  Middle: comparison of different fitting forms of
  mass accretion history: exponential model (Equation (\ref{eq:MAH-1})) with
  one free parameter $\alpha$;
  the exponential-plus-power law model with two free
  parameters $\beta$ and $\gamma$ (Equation (\ref{eq:MAH-2})) or one free parameter
  (Equation (\ref{eq:MAH-3})).  The one-parameter model using an
  optimal relation between $\beta$ and $\gamma$ (the blue
  curve; also see the inset) 
  provides a compelling description of the mass accretion history.
  Right: the cumulative distribution of several proxies for formation
  time: $\zhalf$ (the exact half-mass redshift);
  $z_{\alpha}= \ln2 / \alpha$;
  $z_{\gamma-\beta}= \ln2/(\gamma-\beta)$;
  $z_{\beta\gamma}$ (half-mass redshift obtained by solving Equation (\ref{eq:MAH-2})).
  These different formation time proxies probe somewhat different epochs in a halo's
  history.
}
\label{fig:MAH_app}
\end{figure*}

For each mass accretion history model described in Section \ref{sec:MAH}, we minimize the target
function
\beq
\Delta_{\rm model}^2 = \frac{1}{N} \sum_{i=1}^{N} \left[\log_{10} {M(z_i)}- \log_{10} 
{M_{\rm model}(z_i)}\right]^2.
\label{eq:MAH_res}
\eeq
We note that this function differs from what was used in \cite{McBride09}
\beq
\Delta_{\rm model}^2 = \frac{1}{N} \sum_{i=1}^{N} 
\frac{\left[M(z_i)/\Mtoday - M_{\rm model}(z_i)/\Mtoday \right]^2}{M(z_i)/\Mtoday}.
\eeq
Because our mass accretion history spans approximately 5 orders of
magnitude in mass and starts from redshift 12, weighting by
$M(z_i)/\Mtoday$  significantly underweights high redshift outputs.

The left panel of Figure~\ref{fig:MAH_app} shows the mass growth rate
defined as \beq -\frac{d\ln \Mvir}{dz} \eeq as a function of $z$.  In
Section \ref{sec:MAH} we have discussed the deviation of mass accretion
history from a power law, and this figure clearly shows that the mass
accretion rate is not constant and presents a large amount of scatter.

The middle panel presents the cumulative distribution of rms
residuals, $\Delta_{\rm model}$ (Equation (\ref{eq:MAH_res})), for the fits
to various models of accretion history.  The red curve
corresponds to the exponential model, which has the largest residuals;
the green curve corresponds to the exponential-plus-power law model
with two free parameters ($\beta$, $\gamma$), which has the smallest
residuals.  In between these two models, we compare two 1-parameter
models: the purple curve corresponds to the model that uses the linear
fit between $\beta$ and $\gamma$ ($\beta = 4.16\gamma- 4.00$) to
eliminate one parameter; the blue curve corresponds to an
optimization of the relation between $\beta$ and $\gamma$ to minimize
the overall residuals, obtained with several iterations.  The optimal
relation is given by
\beq
\beta = 5.27 \gamma - 4.61
\eeq
This 1-parameter model performs almost equally well as the 2-parameter
model does.

The right panel of Figure~\ref{fig:MAH_app} shows the cumulative distribution
functions for several halo formation time proxies: 
\begin{itemize}
\item $\zhalf$, the redshift when the halo first reaches half
of its final mass.
\item $z_{\alpha}= \ln 2 / \alpha$, the redshift at which $M(z_{\alpha})=\Mtoday/2$ for the
exponential fit.
\item $z_{\gamma-\beta} = \ln 2 /(\gamma-\beta)$, analogous to $z_\alpha$, where
 $\beta$ and $\gamma$ come from a fit to the exponential-plus-power law model (Equation (\ref{eq:MAH-2})).
 We note that $z_{\gamma-\beta}$ is equivalent to the value of $z_{\alpha}$ measured with only the low redshift outputs.
\item $z_{\beta\gamma}$, obtained by numerically solving
  $M(z_{\beta\gamma})=\Mtoday/2$ using the exponential-plus-power law model (Equation (\ref{eq:MAH-2})).
\end{itemize}

None of the formation time proxies obtained from the fitting functions
captures the exact half-mass redshift.  The rank correlation between
$z_{\alpha}$ and $\zhalf$ is 0.55, and that between $z_{\gamma-\beta}$
and $\zhalf$ is 0.69.  Since we fit for 5 orders of magnitude in mass
growth ($10^{10}$ to $10^{14.8}\hiMsun$), these functions are not
flexible enough to describe the late stage of halo evolution.  
Nevertheless, these
different formation time definitions are still useful because they probe
different epochs in a halo's history; $z_\alpha$ tends to be a
slightly earlier epoch, while $z_{\gamma-\beta}$ and $z_{\beta\gamma}$
tend to be later compared to $\zhalf$.  Although $z_{\beta\gamma}$ is
completely correlated with $z_{\gamma-\beta}$, it corresponds to a
slightly earlier redshift.

\section{Fitting the density profile: goodness of fit}\label{app:concentration}
For each model described in Section \ref{sec:concentration}, we apply the
maximum-likelihood method for fitting
the density profiles of individual halos.
The procedure of finding the maximum-likelihood estimator is to
maximize the log-likelihood function over a set of parameters,
$\mathbf{p}$. The log-likelihood function is defined as
\beq
\ell(\mathbf{p}) = \frac{1}{N} \sum_i \log(\nu_\mathbf{p}(r_i)),
\eeq
where the summation runs over all the $N$ particles, and
\beq
\nu_\mathbf{p}(r) = \frac{1}{M} 4 \pi r^2 \rho_\mathbf{p}(r)
\eeq
so that $\int \nu_\mathbf{p}(r) dr = 1$.
This approach is consistent with the radially-binned fitting method
with a large number of particles
and is more stable than using a fixed number of bins when the halo has fewer particles.

\begin{figure*}[t]
\centering
\includegraphics[width=0.33\columnwidth]{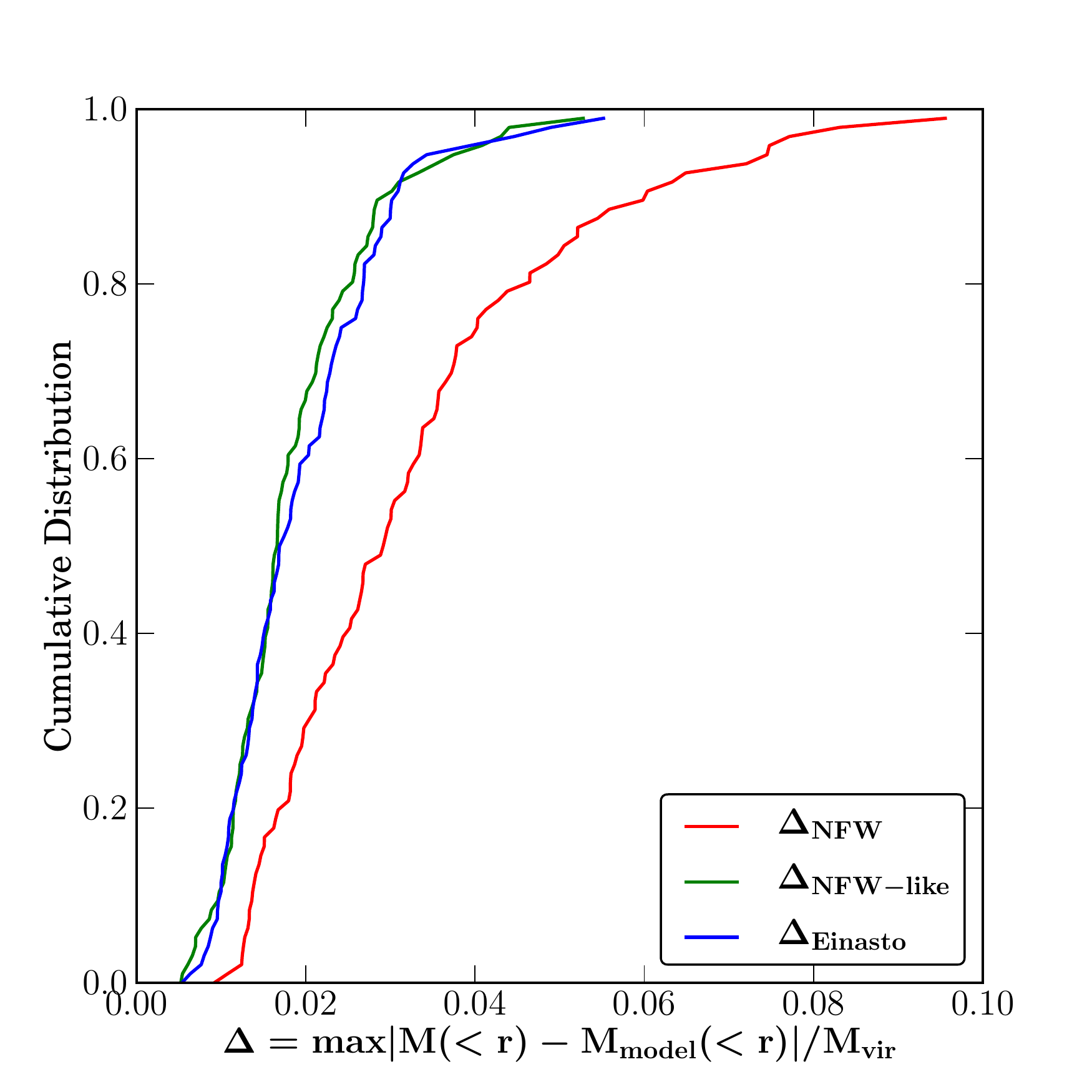}
\includegraphics[width=0.33\columnwidth]{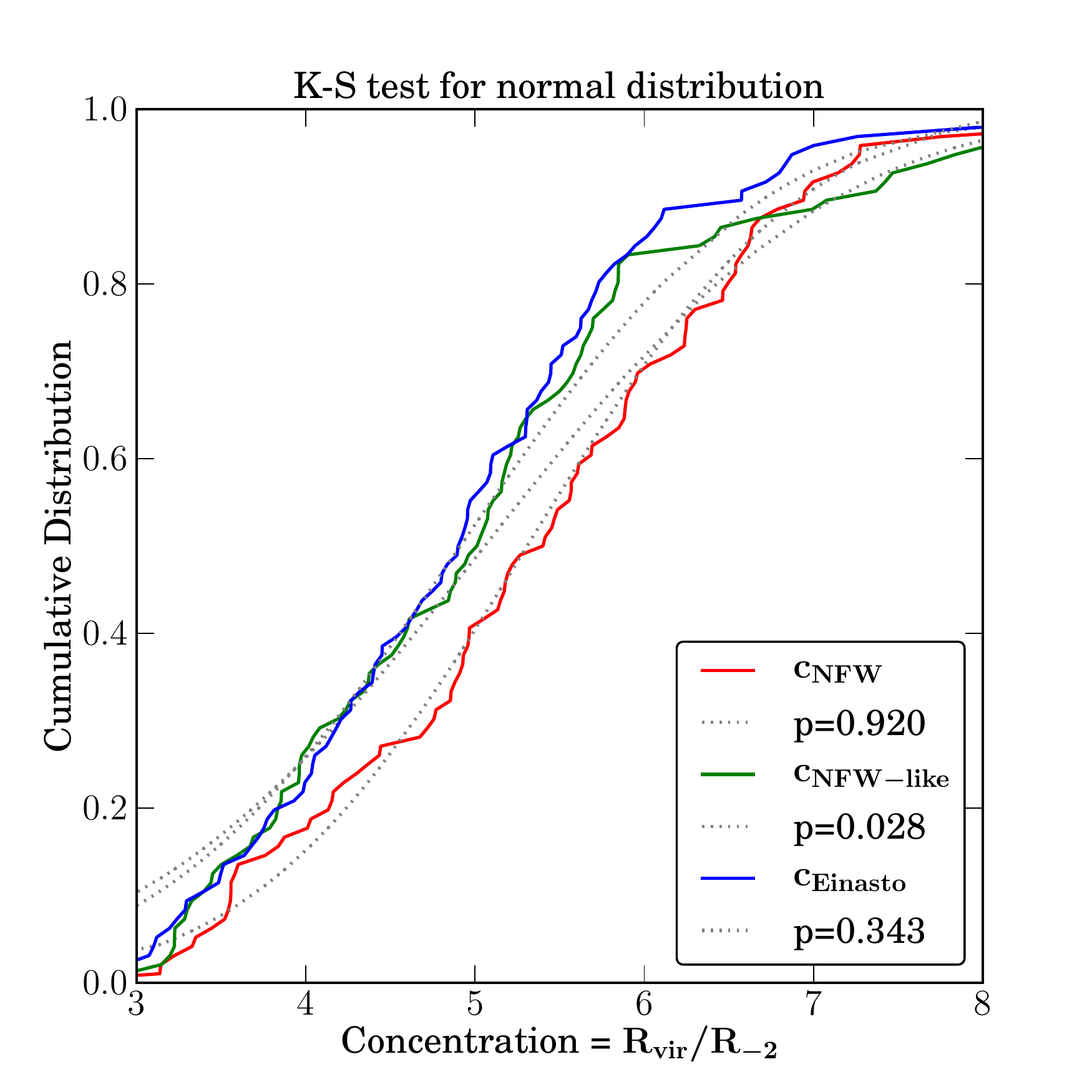}
\includegraphics[width=0.33\columnwidth]{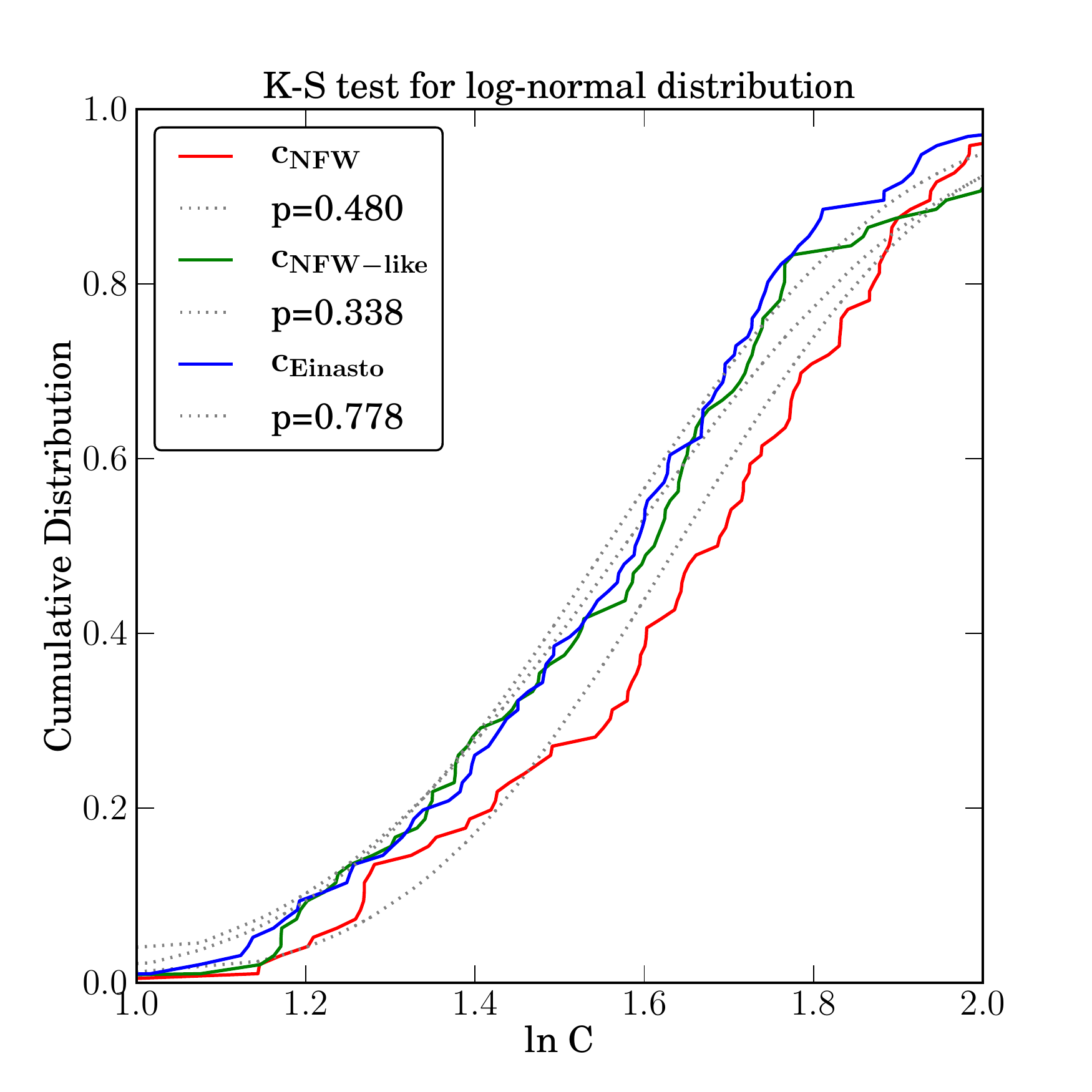}
\caption[]{Left: the Kolmogorov--Smirnov statistics for three
models for the density profile: NFW, NFW-like (with a free outer
slope), and Einasto.  The NFW-like and the Einasto models work equally
well.  Middle and right: concentration distributions based on:
$c_{\rm NFW}$, $c_{\rm NFW-like}$, and $c_{\rm Einasto}=\Rvir/R_{-2}$.
Middle/right shows the cumulative distribution of $c$/$\ln c$
and the corresponding best-fit of normal/log-normal distribution (gray
curve).  Both normal and log-normal can well describe $c_{\rm NFW}$
and $c_{\rm Einasto}$, while $c_{\rm NFW-like}$ prefers a log-normal
distribution.}
\label{fig:concentration_app}
\end{figure*}
The left panel of Figure~\ref{fig:concentration_app} presents the
Kolmogorov-Smirnov statistics for these three models, where
\beq
\Delta_{\rm model} = \frac{{\rm max} | M(<r) -M_{\rm model}(<r)|}{\Mvir}
\eeq
The two-parameter models, the NFW-like and the Einasto models, work
almost equally well and are significantly better than the
single-parameter NFW model.

In the middle/right panel, we show the cumulative distribution for
$c$/$\ln c$ for the three different models stated above.  We also show
the corresponding best fit normal/log-normal distributions and list
the $p$-value based on a Kolmogorov--Smirnov test for goodness of fit.
For the NFW and the Einasto models, both normal and log-normal
distribution provide acceptable descriptions.  For the NFW-like
profile, a log-normal distribution provides a slightly better
description.

\bibliographystyle{apj}
\bibliography{master_refs}
\end{document}